\documentclass[letter,12pt]{article}

\usepackage[dvips]{graphicx}
\usepackage{psfrag}
\usepackage{color}
\usepackage{amsmath}

\newcommand{\sq}{$^2$}

\newcommand{\ssttot}{\sin^2(2\theta_{13})}
\newcommand{\ssttotw}{\sin^2(2\theta_{12})}
\newcommand{\sstttt}{\sin^2(2\theta_{23})}

\author{Fritz DeJongh\footnote{fritzd@fnal.gov}\\ \textsl{Fermilab}} 

\title{
\begin{flushright}
\normalsize
FERMILAB-Pub-02/033-E  
\end{flushright}
\vskip 2ex
Long Baseline Neutrino Physics: \\ From Fermilab to Kamioka}

\begin{document}

\maketitle

\begin{abstract}
  We have investigated the physics potential of very long baseline
  experiments designed to measure $\nu_\mu \to \nu_e$ oscillation
  probabilities.  The principles of our design are to tune the beam
  spectrum to the resonance energy for the matter effect, and to have
  the spectrum cut off rapidly above this energy.  The matter effect
  amplifies the signal, and the cut-off suppresses backgrounds which
  feed-down from higher energy.  The signal-to-noise ratio is
  potentially better than for any other conventional $\nu_\mu$ beam
  experiment.
  
  We find that a beam from Fermilab aimed at the Super-K detector has
  excellent sensitivity to $\ssttot$ and the sign of $\Delta M^2$.  If
  the mass hierarchy is inverted, the beam can be run in antineutrino
  mode with a similar signal-to-noise ratio, and event rate 55\% as
  high as for the neutrino mode.
  
  Combining the Fermilab beam with the JHF-Kamioka proposal adds very
  complementary information.  We find good sensitivity to maximal CP
  violation for values of $\ssttot$ ranging from 0.001 to 0.05.
\end{abstract}

\section{Introduction}
\label{sec:intro}

Measurements of the atmospheric neutrino flux with the
Super-Kamiokande detector have shown that the deficit of muon
neutrinos depends on the zenith angle as expected for neutrino
oscillations~\cite{Fukuda:1998mi}.  This result combined with many
other experimental constraints provides strong evidence for $\nu_\mu
\to \nu_\tau$ oscillations, with mixing $>0.88$\% (90\% C.L.)  and
$\Delta M^2$ in the range $(1 - 5) \times 10^{-3}$
eV$^2$~\cite{Kajita:zv}.

Efforts to study these oscillations with accelerator-produced
neutrinos are in progress, using $\nu_\mu$ beams produced from pion
decay.  The K2K collaboration has detected neutrino interactions in
the Super-K detector using a neutrino beam from KEK 250 km distant.
They report a deficit of events compared to the expectation with no
oscillations~\cite{Ahn:2001cq}.  The MINOS experiment is under
construction, and will use an Iron/scintillator detector 730 km
distant from a neutrino beam produced at Fermilab~\cite{minos}.  These
experiments are expected to confirm that the $\nu_\mu$ disappearance
has the $L/E$ dependence expected for oscillations, and improve the
precision on $\Delta M^2$.

Measurements of solar neutrino fluxes also provide evidence for neutrino
oscillations, usually explained as $\nu_e \to \nu_\mu$ oscillations with
$\Delta m^2$ much smaller than the $\Delta M^2$ for atmospheric oscillations.
The Large Mixing Angle (LMA) solution is strongly favored by the data,
with $\Delta m^2$ in the range 
$(3 - 25) \times 10^{-5}$ eV$^2$~\cite{Smy:2002fs}.
The KamLAND~\cite{Piepke:tg} 
and Borexino~\cite{Ranucci:tc} experiments, currently under construction,
are expected to test the LMA solution and better constrain the 
mixing parameters.

The atmospheric and solar neutrino oscillations can be simultaneously
explained with mixing in a three-flavor model: The the three known
flavors of neutrinos, $\nu_e$, $\nu_\mu$, and $\nu_\tau$, are mixtures
of mass eigenstates $\nu_1$, $\nu_2$, and $\nu_3$.  The data require a
mass hierarchy, with $|m_3 - m_2| \gg |m_2 - m_1|$, The hierarchy can
be ``normal'', with $m_3 > m_2$, or ``inverted'', with $m_2 > m_3$.
We identify $\Delta M^2 = m_3^2 - m_2^2$ and $\Delta m^2 = m_2^2 -
m_1^2$.

\begin{figure}[!t]
  \centering \includegraphics[width=.7\textwidth]{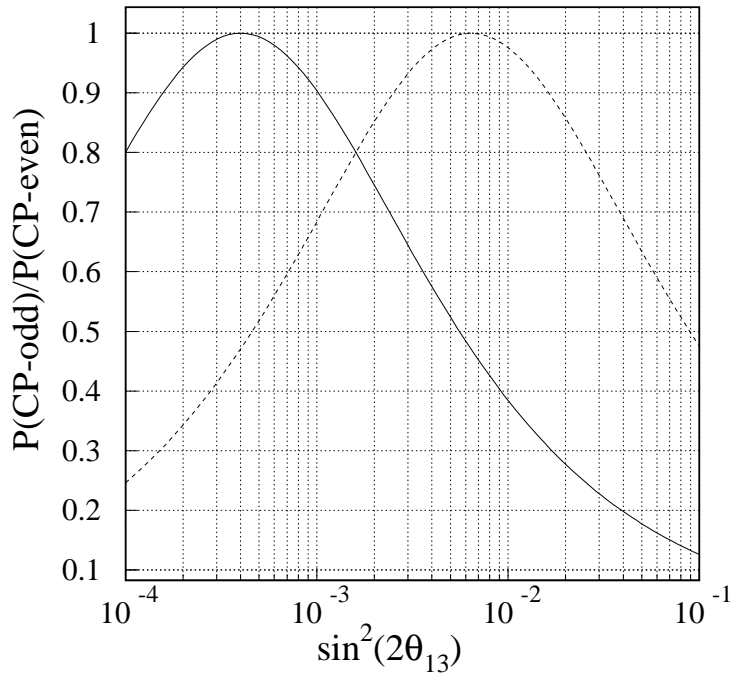}
  \caption{CP asymmetry at the peak of the atmospheric oscillation
    for the parameters of Table~\ref{tab:params}.
    Solid line is for $\Delta m^2 = 5.0\times 10^{-5}$ eV$^2$, and dashed
    line is for $\Delta m^2 = 2.0\times 10^{-4}$ eV$^2$. }
  \label{fig:cp}
\end{figure}

While $\nu_\mu \to \nu_e$ oscillations are mainly associated with
solar oscillations, they can also occur at the atmospheric frequency
if there is a $\nu_3$ component in the $\nu_e$ eigenstate.  Reactor
experiments have placed limits on the size of this mixing, resulting
in the limit $\ssttot < 0.1$~\cite{Apollonio:1999ae,Boehm:2001ik}.  If
$\ssttot > 0$ can be detected, the phenomenology of $\nu_\mu \to
\nu_e$ oscillations becomes very rich:
\begin{itemize}
\item Interference between the solar and atmospheric oscillation amplitudes
  can result in large CP violation~\cite{Mocioiu:2001jy}.
\item Matter effects can modify the oscillation probability, depending
  on the sign of 
$\Delta M^2$~\cite{Mocioiu:1999ag,Mocioiu:2000st,Freund:1999gy}.
\end{itemize}

To study these effects on the oscillation probability, we use the
parametrization of Ref.~\cite{Cervera:2000kp}, with the oscillation
parameters of Table~\ref{tab:params} as defaults.  For example,
Fig.~\ref{fig:cp} shows the possible CP asymmetry as a function of
$\ssttot$, at the peak of the atmospheric oscillations.  While the
mixing in solar oscillations is greater than in atmospheric
oscillations, the solar oscillations develop much more slowly, so the
two amplitudes for $\nu_\mu \to \nu_e$ can be comparable at
atmospheric baselines.  We note that maximal CP asymmetry is a
possibility: There could be a large signal for neutrinos and no signal
at all for antineutrinos, or vice-versa.

\begin{table}[!t]
  \centering
  \begin{tabular}[c]{l l}
    \hline
    $\ssttot$ & variable \\
    $\Delta M^2$ & $3.5\times 10^{-3}$ eV$^2$ \\
    $\Delta m^2$ & $5.0\times 10^{-5}$ eV$^2$ \\
    $\sstttt$ & 1.0 \\
    $\ssttotw$ & 0.8 \\
    $\delta$ & $\pi$/2 \\
    \hline
  \end{tabular}
  \caption{Default oscillation parameters}
  \label{tab:params}
\end{table}

Given the potential for profound insights into fundamental physics,
most studies of future neutrino oscillation experiments with
conventional $\nu_\mu$ beams have focused on $\nu_\mu \to \nu_e$
oscillations.  A wide variety of neutrino energies and detector
technologies have been considered in
Refs.~\cite{Barger:2001qd,Barger:2000nf,Barger:2001qs,pdstudy}.

The potential of low energy neutrino beams combined with
large water-cerenkov detectors has been explored for beams
based at JHF~\cite{Itow:2001ee}, CERN~\cite{Gomez-Cadenas:2001eu},
and Fermilab~\cite{pdstudy} and baselines of 100 to 300 km.
These low energy approaches have the advantage of very low 
backgrounds, and have excellent sensitivity to very small $\nu_e$
appearance signals.  They have the disadvantage that matter effects
are too small to disentangle from CP violating effects, but are still large
enough that ambiguities from the unknown sign of $\Delta M^2$ affect
the ability of the experiment to establish CP violation~\cite{Barger:2001yr}.

The potential of higher energy neutrino beams with very long baseline
($L > 7000$ km) has been explored in
Refs.~\cite{Barger:2001qd,pdstudy}.  The backgrounds in water cerenkov
detectors are larger for higher neutrino energies.  However, the
$\nu_\mu \to \nu_e$ signal at these baselines can be highly amplified
by matter effects, improving the signal-to-background
ratio~\cite{Mocioiu:1999ag,Mocioiu:2000st,Freund:1999gy}.  Also, the
amplified signal depends mostly on atmospheric oscillations, with
solar oscillations playing a much smaller role than with shorter
baselines.

From this summary of short- and long-baseline neutrino
oscillation experiments, we see that they can play very complimentary
roles.  The short-baseline experiment measures a CP-violating
combination of atmospheric and solar oscillations.  The long-baseline
experiment determines the sign of $\Delta M^2$, constrains the matter
effects, and constrains the size of the atmospheric $\nu_\mu \to \nu_e$
oscillation alone.

We propose herein an optimized long-baseline experiment to complement
a short-baseline experiment.  For concreteness, we assume a target
detector at the Kamioka site in Japan, where the Super-K detector
already exists.  The JHF-Kamioka proposal~\cite{Itow:2001ee} uses the
Super-K detector for the first phase, and for the second phase assumes
construction of a new detector with 40 times more fiducial mass.  We
assume that this detector will be used as the target for two neutrino
beams: One from JHF, with a baseline of 295 km, and one from Fermilab,
with a baseline of 9300 km.  The concept of our proposal does not depend on
the details of this choice, and could be adapted
for other locations.  The main principles of our proposal are:
\begin{enumerate}
\item Combine information from very short and very long baselines in order
  to determine the mass hierarchy and CP-violating oscillation parameters.
\item Match the spectrum of the long-baseline neutrino beam to the energy
  at which matter effects produce the maximum amplification of the signal.
\item Design the long-baseline
  neutrino beam to have an energy spectrum with a rapid
  cut-off above the signal region.  Since most
  backgrounds feed-down from the neutrino energy to a lower visible energy,
  this reduces the background in the signal region.
\end{enumerate}

\section{The $\nu_\mu \to \nu_e$ probability at 9300 km}
\label{sec:prob}

We start in the ``leading approximation'' which parametrizes
oscillations driven by the atmospheric $\Delta M^2$ and neglects
oscillations driven by the solar $\Delta m^2$.  The $\nu_\mu \to
\nu_e$ probability can then be written as
\begin{equation}
P(\nu_\mu \to \nu_e) = R^2_m \sin^2 \theta_{23} \sin^2 2\theta_{13}
\sin^2 \left( \frac{1.267\ \Delta M^2}{R_m} \frac{L}{E_\nu} \right) .
\label{eq:mu_e_prob}
\end{equation}
The effects of interaction with matter are included in the parameter $R_m$.
In vacuum, $R_m = 1$.
For positive $\Delta M^2$, there is a resonant energy at which
there is full mixing, with
$R_m = 1/(\sin \theta_{23} \sin 2\theta_{13})$.  
For antineutrinos, this resonance occurs for negative $\Delta M^2$.
Since we're
considering the case $\sin^2 2\theta_{13} < 0.1$, the oscillation length
becomes very long, and for terrestrial baselines we can expand the
probability to first order in $L$:
\begin{equation}
P(\nu_\mu \to \nu_e) = \sin^2 \theta_{23} \sin^2 2\theta_{13}
\left( 1.267\ \Delta M^2 \frac{L}{E_\nu} \right)^2 .
\label{eq:prob_expanded}
\end{equation}
Since the flux is proportional to $1/L^2$, we find that the number of
expected signal events is independent of distance.  
Since background rates decline with $L^2$,
longer baselines will provide better signal to noise.
For negative $\Delta M^2$, the signal would be suppressed and 
unobservable.

\begin{figure}[!t]
  \begin{minipage}[t]{0.5\textwidth}
    \flushright
    \includegraphics[width=\textwidth]{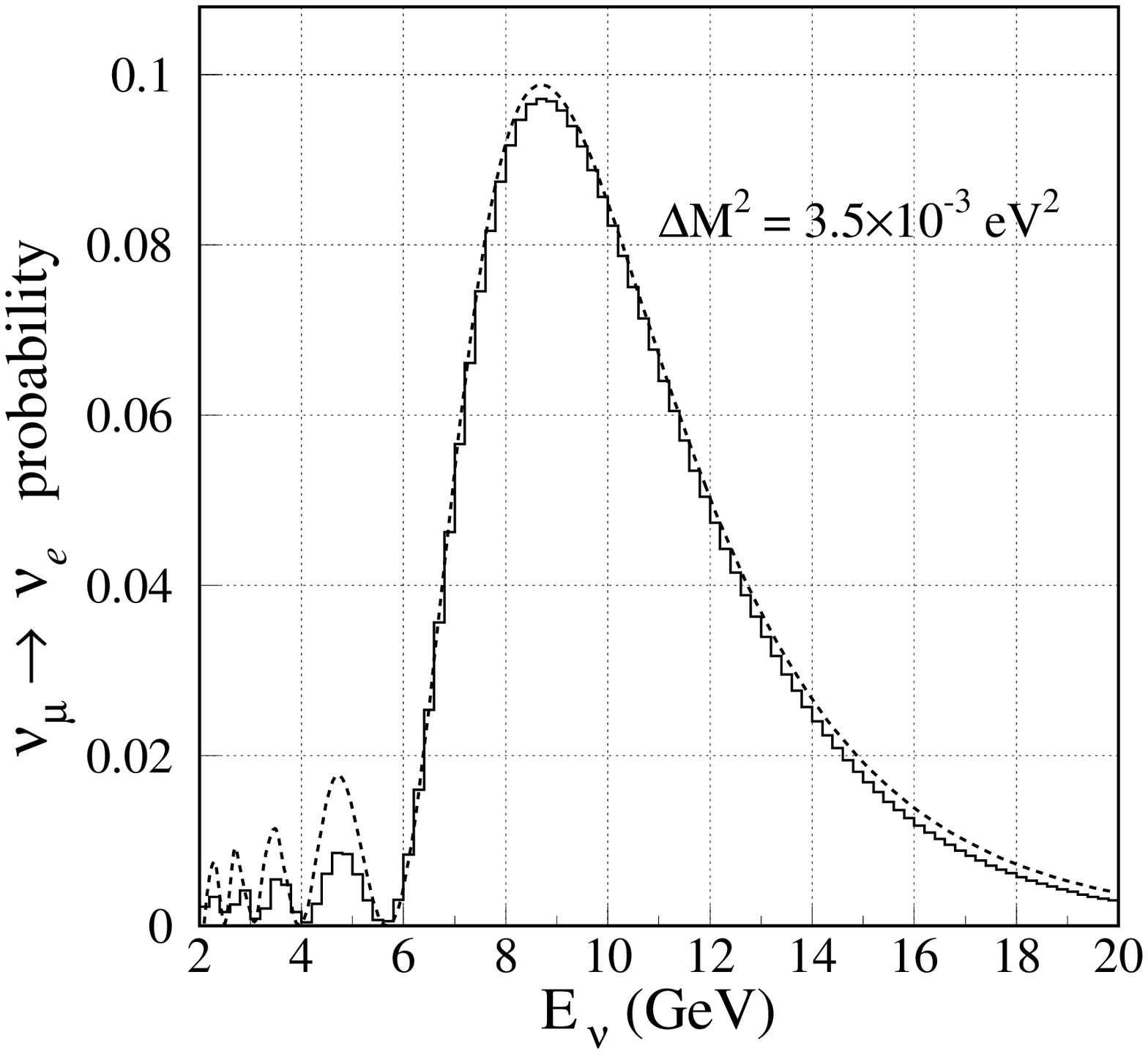}
    \parbox[t]{0.9\textwidth}{
      \caption{Oscillation probability as a function of energy,
        for $L=9300$ km,
        $\ssttot = 0.01$, and neglecting solar oscillations.
        Solid histogram is from a numerical calculation~\cite{stevesplot}, 
        dashed line is from the
        analytical approximation of Ref.~\cite{Cervera:2000kp},
        with the matter parameter $A$ set to 
        $1.75 \times 10^{-4}$ eV\sq/GeV.}
      \label{fig:fkprob}}
  \end{minipage}
  \begin{minipage}[t]{0.5\textwidth}
    \flushright
    \includegraphics[width=\textwidth]{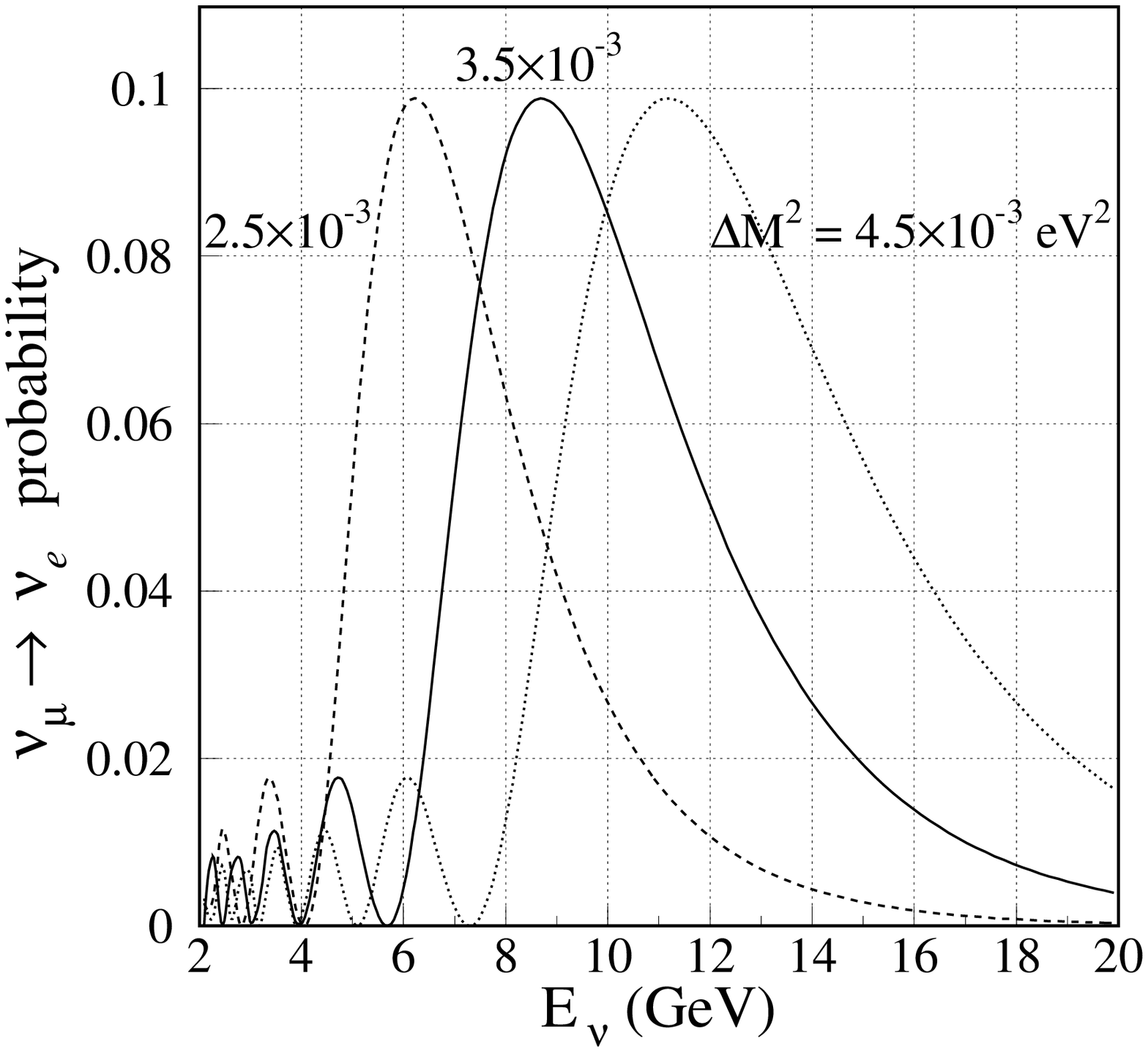}
    \parbox[t]{0.9\textwidth}{
      \caption{Oscillation probability as a function of energy,
      for three different values of the atmospheric $\Delta M^2$
      and parameters of Fig.~\ref{fig:fkprob}.}
    \label{fig:pvsdm2}}
  \end{minipage}
\end{figure}

Fig.~\ref{fig:fkprob} shows $P(\nu_\mu \to \nu_e)$ as a function of
neutrino energy for a baseline of 9300 km, $\Delta M^2 = 3.5 \times
10^{-3}$ eV$^2$, and $\sin^2 2\theta_{13} = 0.01$.  We see that the
resonant energy is in the range 8 to 10 GeV.  At this energy, the
oscillation probability is 20 times the maximum probability occurring
in vacuum. 

Fig.~\ref{fig:fkprob} also compares a numerical
calculation~\cite{stevesplot} of the exact theory incorporating the
density variation of the Earth, with an analytical
approximation~\cite{Cervera:2000kp} using an average density.  The
agreement is excellent in the energy range of interest, and we use the
analytical approximation for the rest of this paper.
Refs.~\cite{Mocioiu:1999ag,Mocioiu:2000st} have also shown that 
using an average density is a good approximation for similar parameters.

Fig.~\ref{fig:pvsdm2} shows 
$P(\nu_\mu \to \nu_e)$ for different values of $\Delta M^2$.  The
resonant energy is proportional to $\Delta M^2$, while the maximum
oscillation probability is constant.

The analytical approximation of Ref.~\cite{Cervera:2000kp} can also
be used to include non-leading terms.  The amplification occurs mainly
for the leading terms, so the oscillation probability for a 9300 km baseline
is dominated by the leading terms much more than for shorter baselines.

\section{The neutrino beam}
\label{sec:beam}

Our starting point for the neutrino beamline design is the NuMI
beamline at Fermilab, which is currently under construction.  In the
NuMI beamline, 120 GeV protons from the Main Injector collide with a
target which is designed for an average power of 0.4 MW.  Pions
produced in the target are focused and charge-selected in a double
horn system, and then decay in a 675~m decay tunnel.  This produces a
beam consisting of mainly $\nu_\mu$.  Different configurations of the
target and horns produce spectra peaking at different energies.  There
is a small $\nu_e$ contamination arising from the decay of kaons produced
in the target, and the decay of muons produced in pion decays.  The
level of this contamination varies from 0.6 to 1\%, depending on the
configuration.

The accelerator used to inject protons into the Main Injector is the
Booster ring, which accelerates protons up to 8~GeV.  Proton drivers
with power of 1 MW or more are currently being
designed~\cite{fermilab_pd,linac_pd}.  With a proton driver replacing
the Booster, the Main Injector would be capable of delivering 2~MW of
proton power to target.  In the case of the linac option for the
proton driver, it may be possible to deliver this power at any proton
energy less than 120~GeV~\cite{linac_pd}.

To aim a neutrino beam from Fermilab to the Kamioka site, the
decay tunnel must be angled downwards $46^\circ$ from the horizontal.
At the Fermilab site, the decay tunnel must fit within a 200~m vertical
depth in order to be within rock favorable for construction and to avoid
a water aquifer.  This limits the length of the decay tunnel, and reduces
the efficiency for pions to decay and produce neutrinos.  The length is
further reduced by the vertical space needed to bend the protons down
to the target, and the space needed for the target and focusing, and 
beam dump at the end of the tunnel.

As discussed in Sect.~\ref{sec:backgrounds}, most backgrounds for our
measurement feed-down from the neutrino energy to a lower visible
energy, while for the signal the entire neutrino energy is visible.
Backgrounds in the signal region tend to arise from neutrinos with
higher energy.  Therefore, our goal for the neutrino beamline is to
produce a spectrum which peaks at the energy for the maximum
oscillation probability in Fig.~\ref{fig:pvsdm2}, and falls off
quickly at higher energy.  A saw-tooth shape would be ideal.  The lack
of a high-energy tail might also simplify the radiation shielding
requirements relative to NuMI.

The beamline will need to be tuned for the actual value of $\Delta M^2$
when it is better known;  for now we assume 
$\Delta M^2 = 3.5 \times 10^{-3}$ eV$^2$, and aim for a peak energy 
of 9 GeV.  

\begin{figure}[!t]
  \centering \includegraphics[width=.7\textwidth]{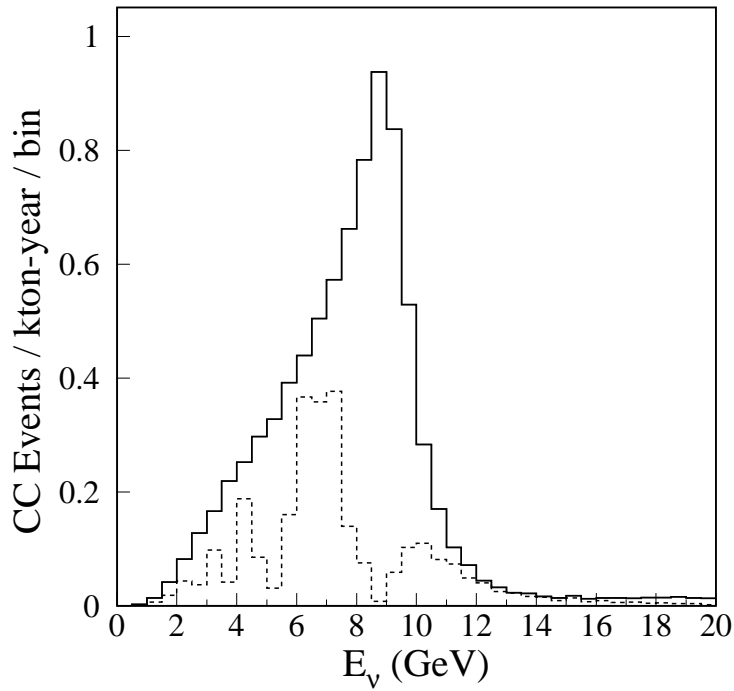}
   \caption{Solid line:  Spectrum of interacting $\nu_\mu$ CC events
     at the Kamioka site, assuming no oscillations.  Dashed line includes
     effects of $\nu_\mu$ disappearance from oscillations.  We note that
     the second disappearance peak happens to occur at the peak of the 
     spectrum.}
   \label{numu_spectrum}
 \end{figure}
 
We have used the NuMI fast beamline simulation to produce a candidate
beamline design.  For the following, we apply a scaling factor of 1.25
needed to bring the fast simulation into agreement with the full simulation
of the medium- and high-energy NuMI configurations~\cite{hylen}.
Starting with a standard NuMI configuration, we have
made the following changes:
\begin{enumerate}
\item Use 60 GeV protons on target instead of 120 GeV.  This reduces the
  cost of the bending magnets, and allows the bend to occur in less vertical space.
  We assume a field of 6~T for the bending magnets.
\item We assume the accelerator complex is upgraded to provide 2 MW of
  proton power from the Main Injector.
\item Change the target and horn
  configuration to produce a spectrum peaking at 10 GeV.
\item Add a small dipole bend after the horns, to bend 20 GeV pions by
  0.5$^\circ$. This allows the higher energy pions to be filtered out, and
  reduces some of the $\nu_e$ component by filtering out neutral kaons.
\end{enumerate}
The resulting spectrum is shown in Fig.~\ref{numu_spectrum}, and has
the basic properties we were aiming for.  The $\nu_\mu$ spectrum including
oscillation effects is also shown.  In an interesting coincidence, the 2nd
disappearance peak happens to occur at the peak of the spectrum.
The length of the pion decay tunnel is 210~m, and the decay efficiency is
37\% what it would be for a NuMI length tunnel.

The event rates do not depend critically on the specifics of the
design.  For example, if we need to use 120 GeV protons to get the
full beam power, the pion decay efficiency relative to NuMI decreases
only to 35\%.  Or, if we use 2~T bending magnets and 60 GeV protons,
the relative efficiency decreases to 33\%.

Ref.~\cite{Aoki:2001rc} has also presented beamline simulations producing
energy spectra with sawtooth shapes, using quadrupole magnets instead
of horns to focus the pions.

\subsection{Flux normalization}
\label{sec:flux}

Since the use of space for
a near detector would further reduce the efficiency of the
beamline, we have assumed that there will be no near detector and the
flux can be normalized using data in the far detector.  As shown in
Fig.~\ref{numu_spectrum}, we expect a large rate of $\nu_\mu$
charged-current interactions in the far detector, and this can in
principle be used to normalize the flux.  The systematic limitations
of this procedure will have to be studied with a detailed simulation.
It's possible that neutral-current interactions or $\nu_\tau$
charged-current interactions might also be used to help measure the flux.

\subsection{Running with antineutrino beams}
\label{sec:antinu}

The neutrino beamline can be converted into an antineutrino beamline
by reversing the polarity of the focusing horns and bending magnet.
Since the antineutrino cross-section is smaller than the neutrino
cross-section, and $\pi^-$ production in the proton target is
suppressed relative to $\pi^+$ production, event rates in the far
detector are typically 1/3 of those in neutrino beams.  The intrinsic
$\nu_e$ and $\bar{\nu}_e$ backgrounds tend to be worse as well.  These
problems are mitigated by the following effects:
\begin{itemize}
\item As discussed in Sect.~\ref{sect:signal}, the efficiency of reconstructing
  $\bar{\nu}_e$ events is higher than for $\nu_e$ events.  After reconstruction
  and event selection cuts, the antineutrino rate will be about 55\% of
  the neutrino rate.
\item The dipole bend helps remove neutral kaons and $K^+$ in the pion
  decay tunnel, removing the $\nu_e$ component of the beam.  The remaining
  $\bar{\nu}_e$ component will be comparable to the $\nu_e$ component
  of the neutrino beam.
\end{itemize}

\section{The $\nu_e$ signal}
\label{sect:signal}

The $\nu_e$ signal is detected from charged-current (CC) interactions
producing an electron along with a hadronic shower with energy $E_{\rm
  had}$.  The visible energy of the event, $E_{\mathrm{vis}}$, (the
energy not carried away by neutrinos) is equal to the neutrino energy.
The differential cross-section in terms of the variable $y = E_{\rm
  had} / E_\nu$, normalized to 1, can be approximately parametrized
by~\cite{Barger:1999fs}:
\begin{equation}
\frac{d\sigma_\nu}{dy} = \frac{15}{16}\left(1+\frac{(1-y)^2}{5}\right) .
\label{dsdy}
\end{equation}
For antineutrinos, the parametrization is:
\begin{equation}
\frac{d\sigma_{\bar{\nu}}}{dy} = \frac{15}{8}\left(\frac{1}{5}
  +{(1-y)^2}\right) .
\label{dsdy_bar}
\end{equation}

We will model the efficiency of detecting a $\nu_e$ CC event by assuming
the fiducial mass of the detector for the target mass, and by requiring
the production of an electron above some threshold $E$.   The
probability of a $\nu_e$ with energy $E_\nu$ generating an electron
above a threshold $E$ is:
\begin{equation}
p_{cc} = \frac{15}{16} \left( \frac{16}{15} - \frac{E}{E_\nu}
- \frac{1}{15}\left(\frac{E}{E\nu}\right)^3  \right) .
\label{eq.pcc}
\end{equation}
For antineutrinos, the parametrization is:
\begin{equation}
p_{cc} = \frac{15}{8} \left( \frac{8}{15} -  \frac{1}{5}\frac{E}{E_\nu}
- \frac{1}{3}\left(\frac{E}{E\nu}\right)^3  \right) .
\label{eq:pcc-bar}
\end{equation}

\begin{figure}[!t]
  \begin{minipage}[t]{0.5\textwidth}
    \flushright
    \includegraphics[width=\textwidth]{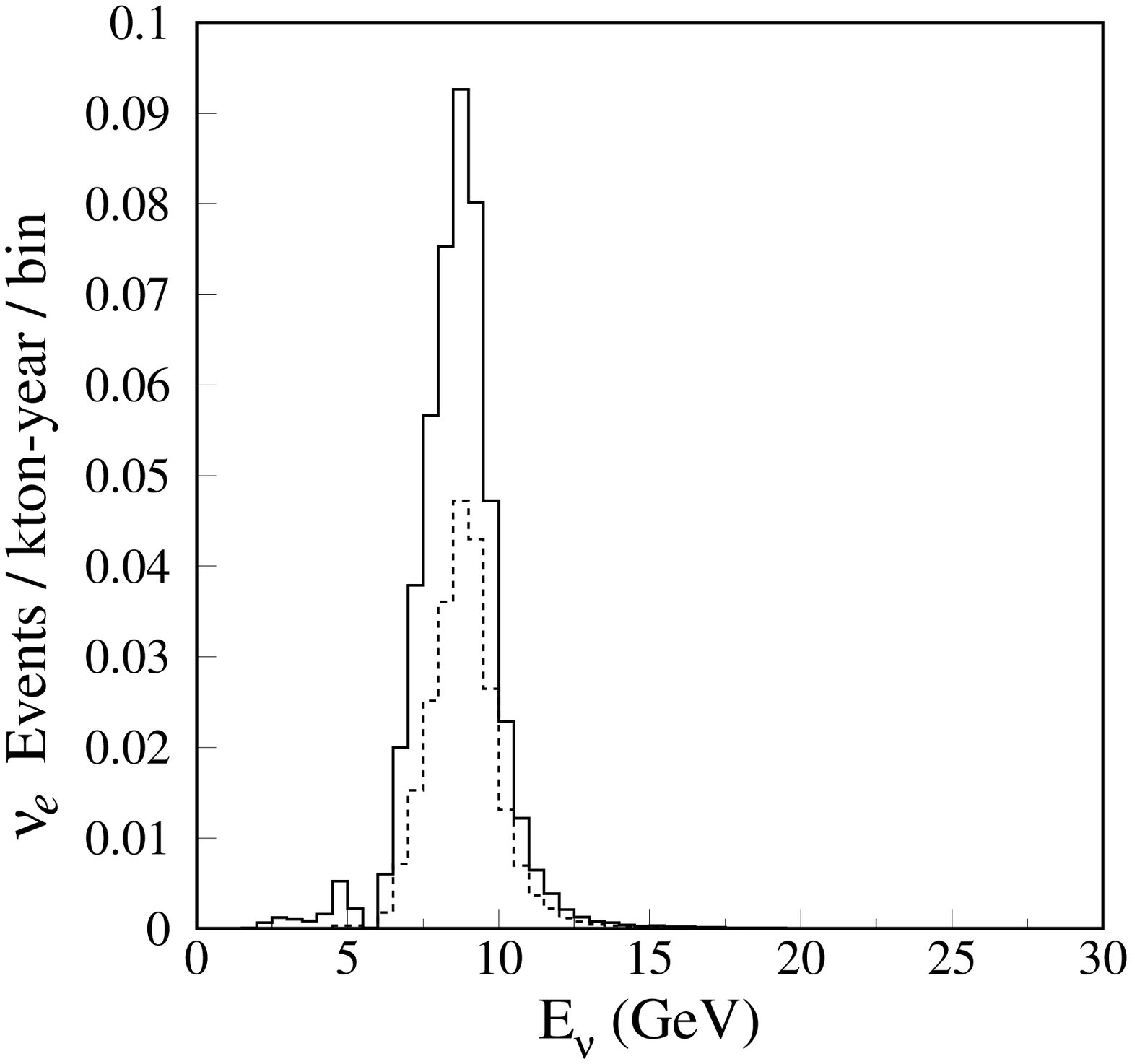}
    \parbox[t]{0.9\textwidth}{
      \caption{Solid line:  Spectrum of interacting $\nu_e$ CC events
        at the Kamioka site, assuming the oscillation
        parameters of Fig.~\ref{fig:fkprob}.  Dashed line includes
        effects of cuts described in Sect.~\ref{sect:signal}.}
      \label{nue_spectrum}}
  \end{minipage}
  \begin{minipage}[t]{0.5\textwidth}
    \flushright
    \includegraphics[width=\textwidth]{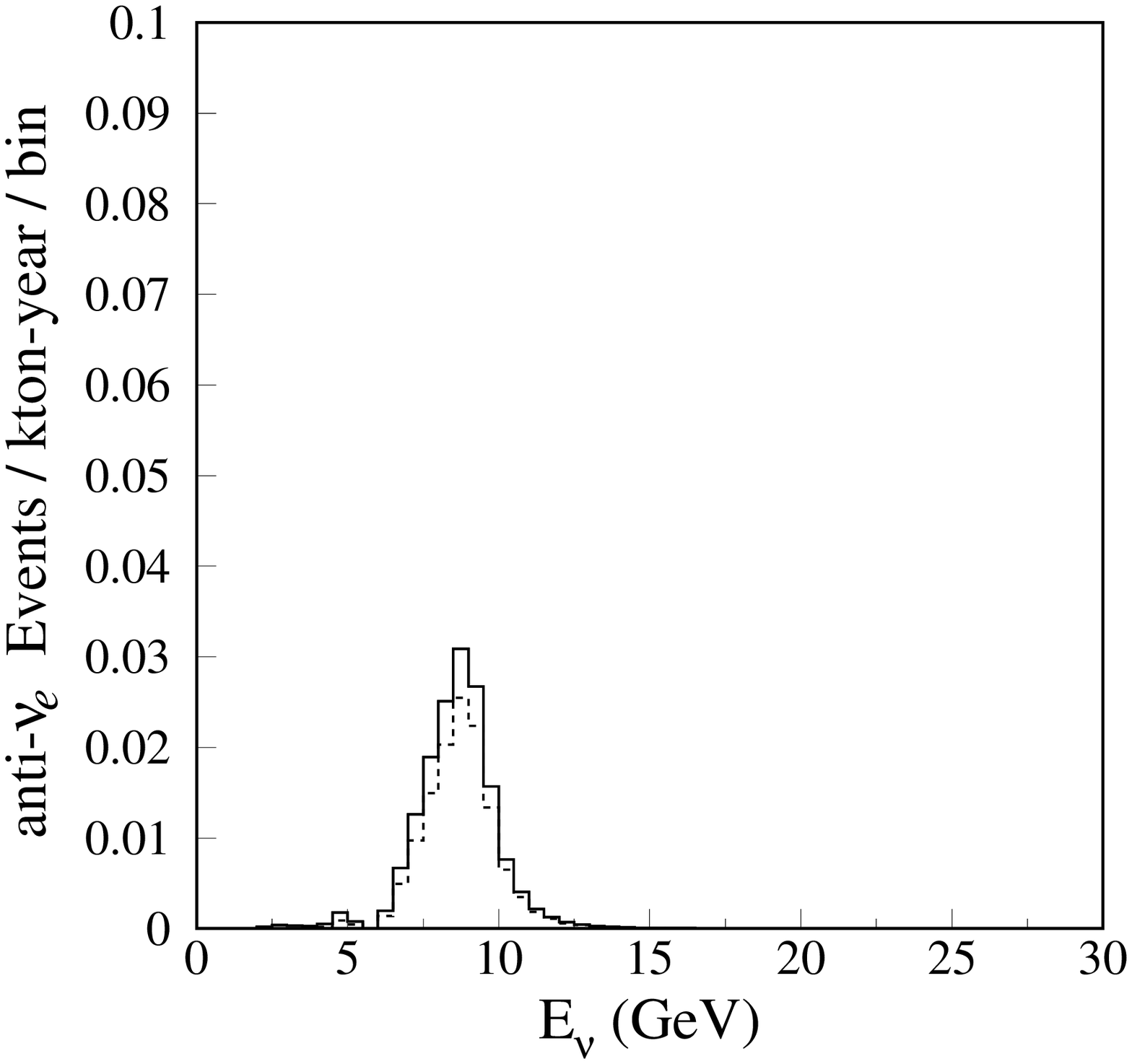}
    \parbox[t]{0.9\textwidth}{
      \caption{Solid line:  Spectrum of interacting $\bar{\nu}_e$ CC events
        at the Kamioka site, assuming the oscillation
        parameters of Fig.~\ref{fig:fkprob}.  Dashed line includes
        effects of cuts described in Sect.~\ref{sect:signal}.
        The rate after cuts is 55\% of the rate for $\nu_e$ CC events.}
      \label{nueb_spectrum}}
  \end{minipage}
\end{figure}
 
For neutrinos, we choose the following cuts for the threshold $E$:
\begin{enumerate}
\item $E > 4.5$ GeV.
\item $E/E_{\mathrm{vis}} > 0.45$.
\end{enumerate}
Fig.~\ref{nue_spectrum} shows the $\nu_e$ CC energy spectrum at 
the Kamioka site, assuming the neutrino beam described in 
Sect.~\ref{sec:beam}, and the neutrino oscillation parameters
of Fig.~\ref{fig:fkprob}.  The dashed line of Fig.~\ref{nue_spectrum}
shows the effect of the above cuts.  The average efficiency of
these cuts is 50\%.  With these cuts, we find 0.21 events
in the signal region,  per kton-year of exposure, where the signal
region is defined as energy between 7 and 10.5 GeV.

The $y$-distribution for antineutrinos is more peaked towards zero
than for neutrinos, as described by Eqs.~(\ref{dsdy}) and (\ref{dsdy_bar}).
Therefore, for the same NC background fraction,
$\bar{\nu}_e$ CC events can be selected with 
more efficient cuts.
For neutrinos, we choose the following cuts for the threshold $E$:
\begin{enumerate}
\item $E > 3.3$ GeV.
\item $E/E_{\mathrm{vis}} > 0.33$.
\end{enumerate}
Fig.~\ref{nueb_spectrum} shows the $\bar{\nu}_e$ CC energy spectrum at 
the Kamioka site, assuming the neutrino beam described in 
Sect.~\ref{sec:beam}, and the neutrino oscillation parameters
of Fig.~\ref{fig:fkprob} with negative $\Delta M^2$.
The dashed line of Fig.~\ref{nueb_spectrum}
shows the effect of the above cuts.  The average efficiency of
these cuts is 80\%, producing an antineutrino signal rate
55\% of the neutrino rate.

\section{Backgrounds}
\label{sec:backgrounds}

\subsection{Backgrounds from neutral-current events}
\label{sec:frag}

Neutral pions are produced in the hadronic showers present 
in all neutrino interactions with nuclei.  At high energies,
the photons from the $\pi^0 \to \gamma \gamma$ decay produce a 
single electromagnetic shower, which in water cerenkov
detectors is indistinguishable from that of an electron.
Neutral-current (NC) events with an energetic $\pi^0$ constitute
a major background to the $\nu_e$ appearance signal.  The visible
energy of these events is given by $y E_\nu$, so they feed-down
from the neutrino energy to a lower visible energy.

We have studied the production of neutral pions with the LEPTO
program~\cite{Ingelman:1996mq} which generates deep inelastic neutrino
interactions and models the fragmentation of the hadronic component.
There are several efforts underway to model resonance and
quasi-elastic production, which we have not included, and which might
account for roughly 15\% of the cross-section in the energy range we
are interested in~\cite{gallagher}.  However, we do expect the
backgrounds to come more from high-$y$ events, or from feed-down from
higher-energy neutrinos, where the fragmentation assumption is more
applicable.

We find that Eq.~(\ref{dsdy}) and
Eq.~(\ref{dsdy_bar})  agree well with LEPTO-generated
data for charged-current and neutral-current events.

\begin{figure}[!t]
  \centering \includegraphics[width=.7\textwidth]{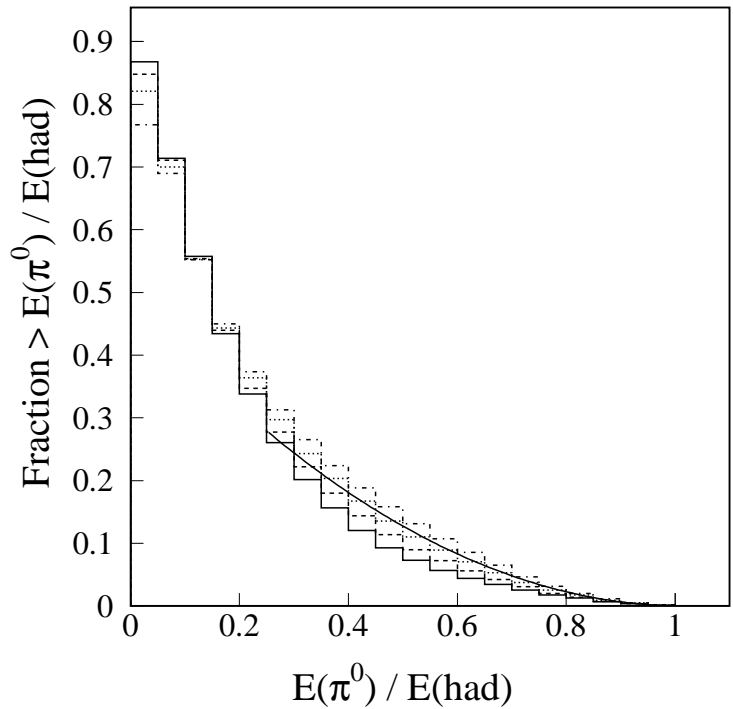}
   \caption{Integral plots of the $E(\pi^0)/E(\rm had)$ distribution in
     neutrino interactions.
     The different histograms are for different ranges of hadronic energy:
     Solid:  20 to 25 GeV.  Dashed:  15 to 20 GeV.  Dotted:  10 to 15 GeV.
     Dot-Dashed:  5 to 10 GeV.  The curve is our parametrization of this
     quantity.}
   \label{pi0frag}
 \end{figure}
 
 The $\pi^0$ fragmentation functions are roughly independent of
 hadronic energy.  This is illustrated in Fig.~\ref{pi0frag}, which
 shows integral plots of the $E(\pi^0)/E(\rm had)$ distribution.  The
probability for a hadronic shower to produce a $\pi^0$ with energy
greater than a fraction $x$ of $E(\rm had)$ can be parametrized by:
\begin{equation}
p(x) = (0.49)-(0.96)x+(0.47)x^2 .
\label{pi0prob}
\end{equation}
For a given neutrino energy, we can use Eq.~\ref{dsdy} to obtain the
visible energy spectrum in NC interactions.  For a given visible energy, we
can then use Eq.~\ref{pi0prob} to calculate the probability that there
will be a $\pi^0$ above some cut-off energy.  Thus, we can convolute an
input neutrino spectrum into an output NC background spectrum.

\begin{figure}[!t]
  \begin{minipage}[t]{0.5\textwidth}
    \flushright
    \includegraphics[width=\textwidth]{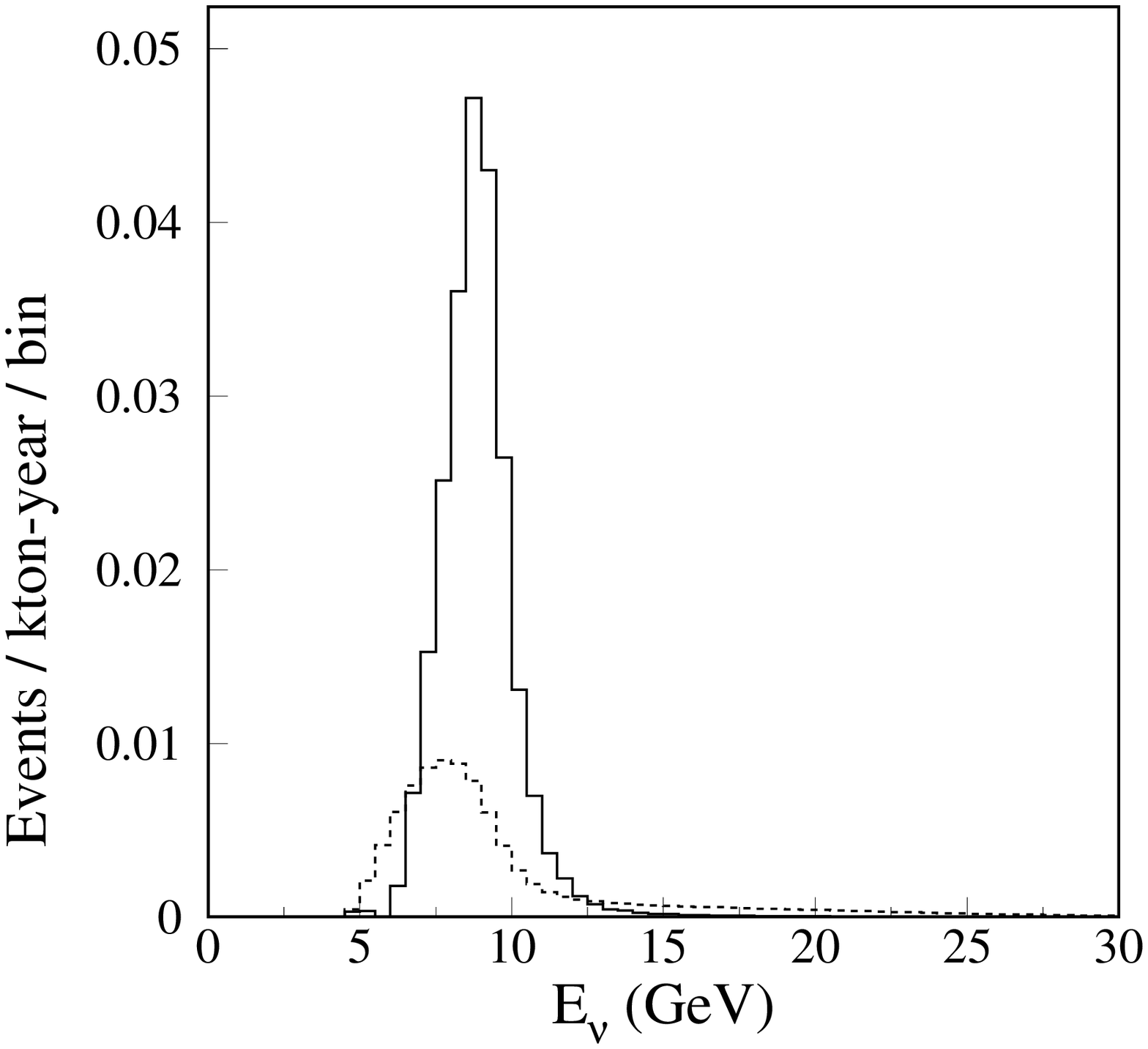}
    \parbox[t]{0.9\textwidth}{
      \caption{Solid line:  Spectrum of $\nu_e$ events with cuts from
        Fig.~\ref{nue_spectrum}.  Dashed line:  Distribution of
        background from NC events.}
      \label{nue_and_nc}}
  \end{minipage}
  \begin{minipage}[t]{0.5\textwidth}
    \flushright
    \includegraphics[width=\textwidth]{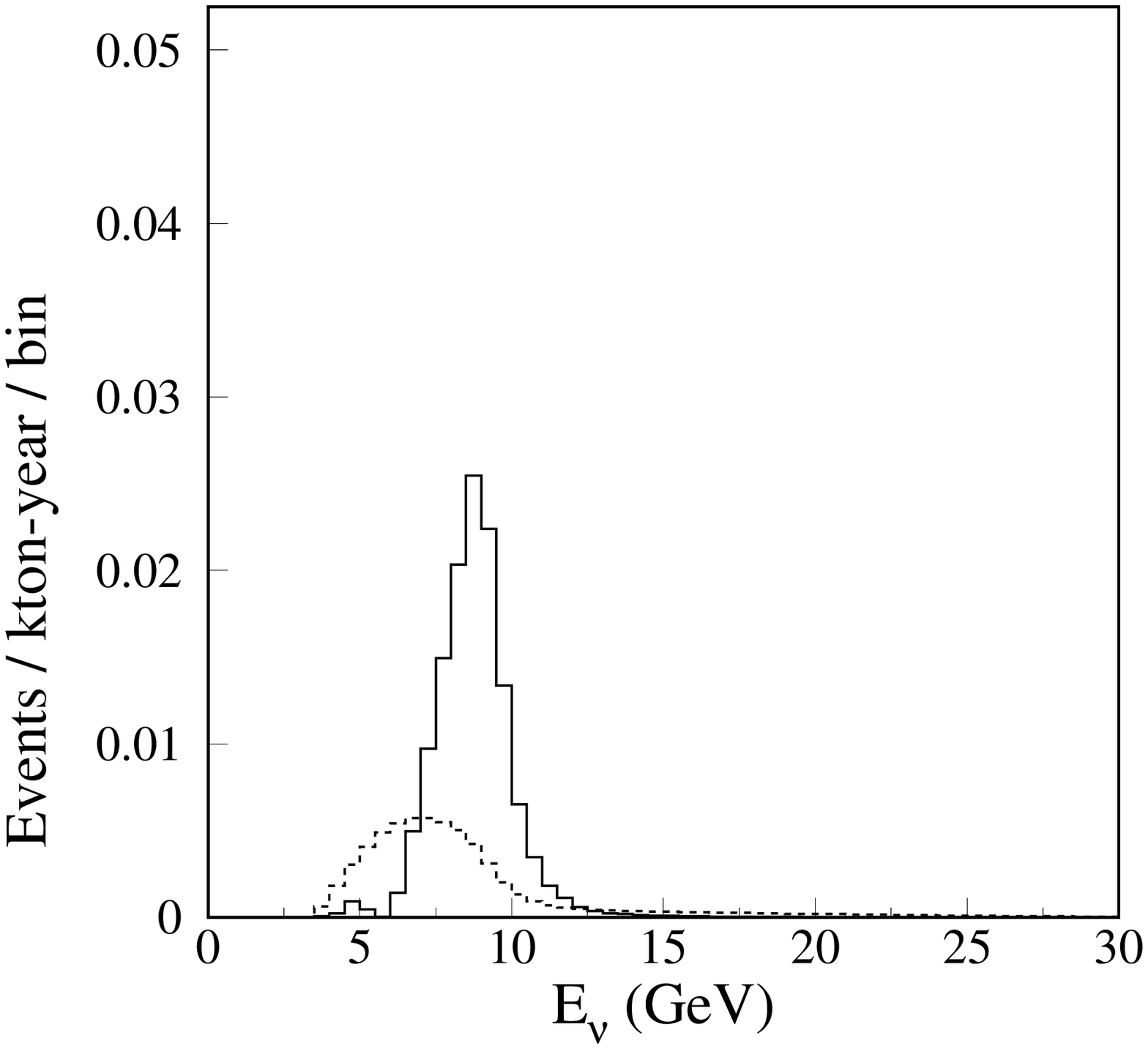}
    \parbox[t]{0.9\textwidth}{
      \caption{Solid line:  Spectrum of $\bar{\nu}_e$ events with cuts from
        Fig.~\ref{nueb_spectrum}.  Dashed line:  Distribution of
        background from NC events.}
      \label{nueb_and_nc}}
  \end{minipage}
\end{figure}

Fig.~\ref{nue_and_nc} shows the NC background 
to the $\nu_e$ signal, estimated with this
procedure with the cuts of Sect.~\ref{sect:signal}.  We find
a background of 0.047 events per kton-year in the signal region, corresponding
to $f_B = 2.1\%$, where $f_B$ is the background fraction relative
to the expected signal after cuts for a beam with 100\% 
$\nu_e$~\cite{Barger:2001qd}.
Fig.~\ref{nueb_and_nc} shows the same distributions for the
$\bar{\nu}_e$ signal.  We obtain the same background fraction with
a signal rate 55\% of the $\nu_e$ rate.

Since the direction of the incoming neutrino is known, and the outgoing
neutrino can carry away transverse energy, the direction of the observed
energy flow might also be used to further reduce NC background.  A detailed
simulation will be needed to investigate this possibility.

\subsection{Background from $\nu_\mu$ charged-current interactions}
\label{sec:numu_cc}

In this analysis, we assume that all $\nu_\mu$ CC events
can be rejected due to the presence of a muon, and only consider the
$\pi^0$ background from NC events.  For high-$y$ events however,
the muon will have low energy and can easily be missed, while
the hadronic shower has most of the energy of the neutrino.
Neutral pions from these hadronic showers will generate backgrounds
the same as for neutral currents.  Fortunately, as illustrated in
Fig.~\ref{numu_spectrum}, the $\nu_\mu$ flux is very low
in the signal region.  This is true for other values of $\Delta M^2$
as well.  So, we expect this source of background to be negligible
compared to the background from neutral current events.

\subsection{Background from $\nu_\tau$ charged-current interactions}
\label{sec:nutau_cc}

At the peak energy for the $\nu_e$ signal, most of the $\nu_\mu$ have
oscillated into $\nu_\tau$, so $\nu_\tau$ CC interactions are a
potentially serious source of background.  However, at this energy the
$\nu_\tau$ CC cross-section is only $\approx$~8\% of the $\nu_e$ CC
cross-section.  The decay mode $\tau \to e \nu \bar{\nu}$ produces
a real electron background, but the branching ratio  is only
18\%.  Neutral pions in $\nu_\tau$ CC events can also produce backgrounds.
Since all these events have an escaping neutrino from the $\tau$ decay,
these backgrounds feed-down to lower visible energy.  As long as we
keep the high energy tail of the neutrino beam suppressed, we expect
this source of background to be small.
However, the cuts needed to suppress this background need more study
with a detailed simulation.

\subsection{Intrinsic $\nu_e$ backgrounds}
\label{sec:beam_nue}

As discussed in Sect.~\ref{sec:beam} the neutrino beam has
a small amount of $\nu_e$ contamination.  The dipole bend in the beamline
can help reduce this background by eliminating neutral kaons in the
beamline.  Also, the $\nu_e$ energy spectrum is very broad, while
the signal region is narrow.  
The intrinsic $\nu_e$ background as a function of the energy width of
the beam has been studied in Ref.~\cite{Barger:2001qd}, showing that it
is greatly reduced for narrow beams.
Therefore, we expect this background also
to be very small.

\section{Comparison with the JHF-Kamioka project}
\label{sec:comp}

The first stage of the JHF-Kamoika project uses a 0.75 MW proton driver
to generate a neutrino beam aimed at the Super-Kamiokande detector.
In phase 2, the proton power is increased to 4 MW, and a new 
water cerenkov detector, called Hyper-K, is built with 40 times more
fiducial mass.

With very similar mixing parameters as those used in 
Sect.~\ref{sect:signal}, the
JHF-Kamioka proposal obtains 0.11 events per kton-year in phase 1,
compared to 0.21 for our proposal.  The event rate for the Fermilab beam
is 200\% of the rate for the phase 1 JHF beam, and 40\% of the rate
for the phase 2 JHF beam.

The JHF-Kamioka project achieves a very low background rate:
$f_B = 0.51\%$, compared to $f_B = 2.5\%$ for the Fermilab beam,
where $f_B$ is the background fraction relative
to the expected signal after cuts for a beam with 100\% 
$\nu_e$~\cite{Barger:2001qd}.
However, the signal from the Fermilab beam is amplified by a factor
of 20, so the effective background for comparison with the JHF beam
is only 0.13\%.   

The signal region for the Fermilab beam is fairly narrow, between
7 and 10.5 GeV.  The signal region for the JHF beam is defined
as energy between 0.4 and 1.2 GeV.  The average oscillation probability
over the JHF spectrum is only 56\% of the peak probability.  This further
accentuates the background advantage of the Fermilab beam.

\section{Combined analysis of data from Fermilab and JHF beams}
\label{sec:combined}

In the previous section, we showed that the Fermilab and JHF beams
would have comparable performance.  
The JHF phase 2 beam has an advantage in rate, while the
Fermilab beam has an advantage in background.
However, the real advantage
comes in combining information from the two beams.  The first
step will be the detection of $\nu_\mu \to \nu_e$ oscillations and 
determination of the sign of $\Delta M^2$.  The second step will
be constraints on sub-leading effects such as CP violation.

\subsection{Detecting $\nu_\mu \to \nu_e$ oscillations and the 
sign of $\Delta M^2$}
\label{sec:detect}

\begin{figure}[!t]
  \begin{minipage}[t]{0.5\textwidth}
    \flushright
    \includegraphics[width=\textwidth]{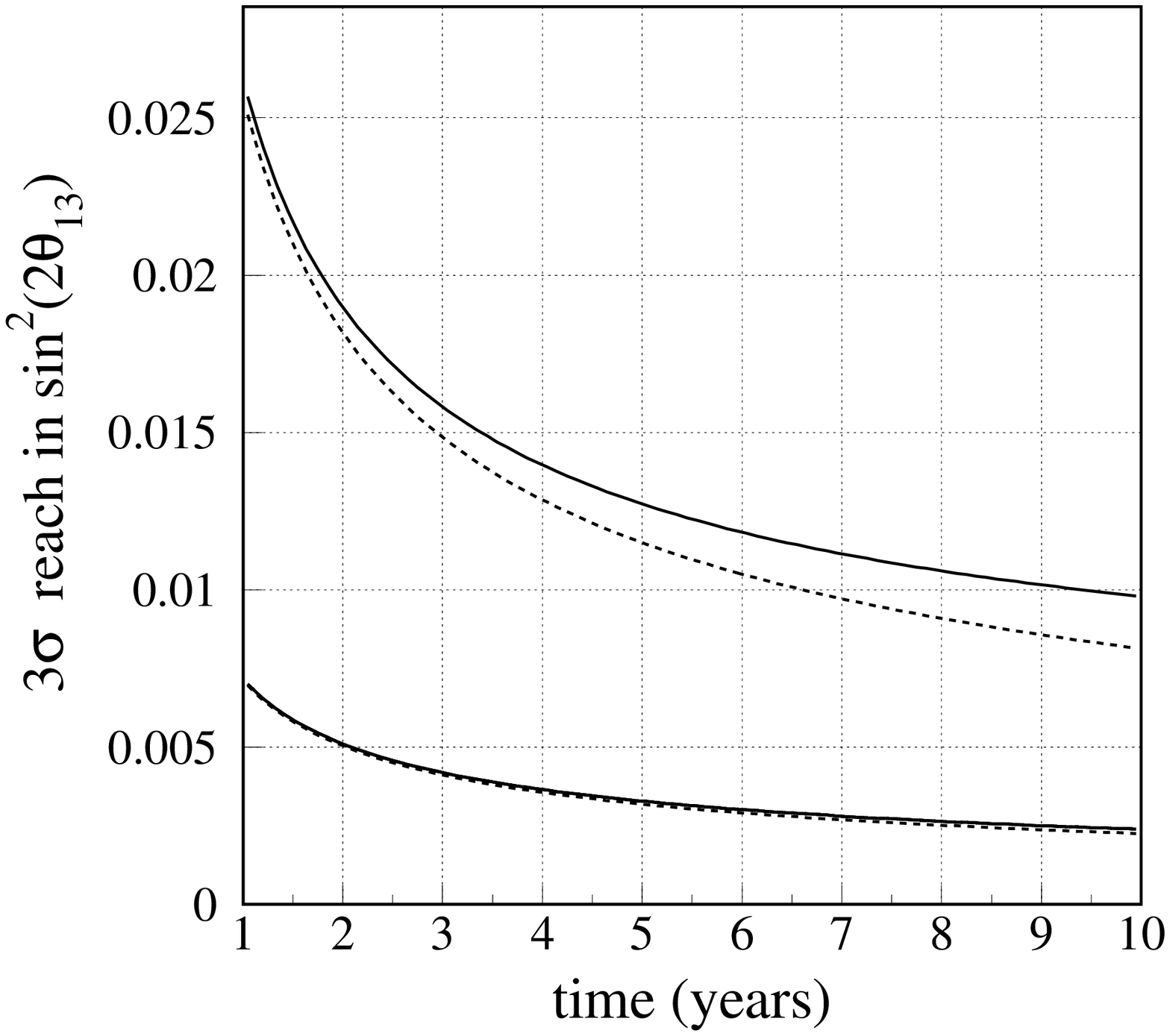}
    \parbox[t]{0.9\textwidth}{
      \caption{Reach in $\ssttot$ with Super-K detector.
        Upper curves are for JHF phase 1 beam, lower curves for Fermilab beam.
        Solid curves assume 10\% systematic uncertainty on the background,
        dashed curves assume no systematic uncertainty.  We assume positive
        $\Delta M^2$, and atmospheric oscillations only.}
      \label{fig:t13_reach_1}}
  \end{minipage}
  \begin{minipage}[t]{0.5\textwidth}
    \flushright
    \includegraphics[width=\textwidth]{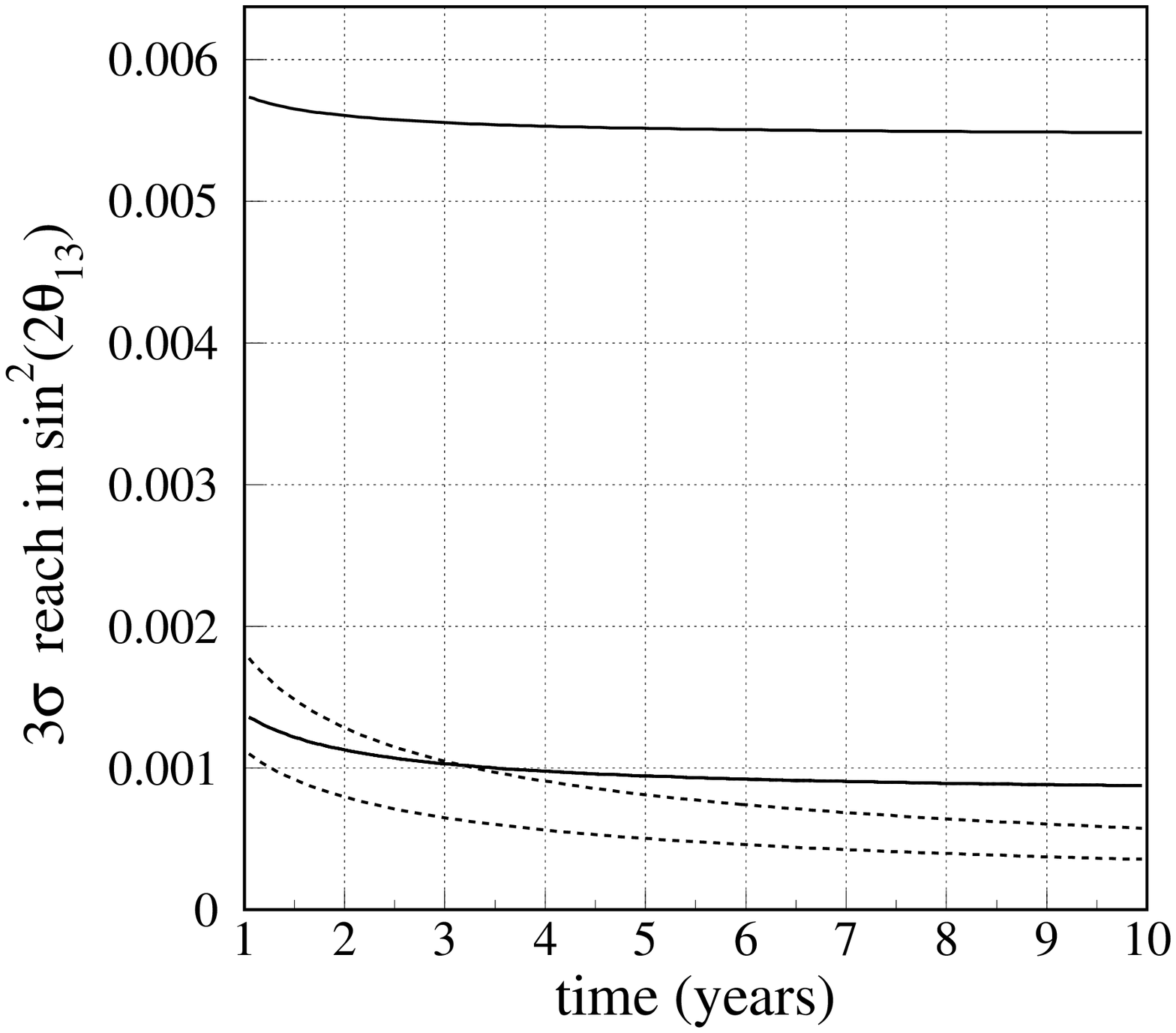}
    \parbox[t]{0.9\textwidth}{
      \caption{Reach in $\ssttot$ with Hyper-K detector.
        Upper curves are for JHF phase 2 beam, lower curves for Fermilab beam.
        Solid curves assume 10\% systematic uncertainty on the background,
        dashed curves assume no systematic uncertainty.  We assume positive
        $\Delta M^2$, and atmospheric oscillations only.}
    \label{fig:t13_reach_2}}
  \end{minipage}
\end{figure}

Fig.~\ref{fig:t13_reach_1} shows
our estimates of the reach in $\ssttot$ for JHF
and Fermilab beams with the Super-K detector, assuming positive
$\Delta M^2$, and atmospheric oscillations only.  
The JHF curves are in reasonable agreement with curves in the
JHF-Kamioka proposal~\cite{Itow:2001ee}.
If a signal appears
with the JHF beam, it should be confirmed by the Fermilab beam within
one year of running, otherwise $\Delta M^2$ must be negative.  It should
then be possible to find the signal with the Fermilab beam in antineutrino mode.
The performance of the Fermilab beam is excellent:   5 years of running 
yields a 3$\sigma$ reach of 0.003 in $\ssttot$.

Fig.~\ref{fig:t13_reach_2} shows a similar plot for the Hyper-K detector.
A 5 year run should produce a 3$\sigma$ reach of 0.001 or better, 
with a strong dependence on the background level and systematic
uncertainty.

It is interesting that if $\ssttot$ is small enough for 
Eq.~(\ref{eq:prob_expanded}) to be valid, the long-baseline
oscillation probability does not depend explicitly on the size of the matter effect
for neutrinos at the resonant energy.  If the peak energy for the matter
resonance can be determined, the theory of matter effects can be tested
and systematic uncertainties from matter effects can be kept 
small~\cite{Freund:1999gy}.

\subsection{Constraints on the CP violating phase}
\label{sec:cp}

The JHF-Kamioka phase~2
program assumes 2 years of neutrino running, and 6
years of antineutrino running.  They have assumed the background will
be very well constrained, and for our analysis we have assumed a 2\%
systematic uncertainty on the background.  Except for the highest
allowed values of $\ssttot$, it will be necessary to either find a way to
further reduce the background, or to determine the background to this
level.  For the Fermilab beam, we have assumed a 5\% systematic
uncertainty on the background, and an 8 year run.

We have neglected the systematic uncertainty on the flux prediction, 
but note that for the highest allowed values of $\ssttot$, 
it will need to be constrained to 1\%.

For the JHF beam,
we have assumed the signal to background ratio in the antineutrino beam
is the same as for the neutrino beam, and the 6 year exposure for the
antineutrino beam is equivalent to 2 years with the neutrino beam.  
Without a dipole bend, it is likely that the backgrounds with the antineutrino
beam will in fact be much worse.  As an alternative, we have considered
running for 8 years with the JHF neutrino beam.

We have assumed the parameter values in Table~\ref{tab:params},
and $\ssttot$  $=$ 0.01, 0.05, or 0.001.  We have not included the effect
of uncertainties in parameters other than $\ssttot$ and $\delta$.

We have assumed that
the JHF beam measures oscillation probabilities at 0.8 GeV, and the
Fermilab beam at 9.0 GeV, although in reality a more sophisticated
energy-dependent analysis will be performed.  The expected oscillation
probability measurements are shown in Table~\ref{tab:measurements}.

\begin{table}[!t]
  \centering
  \begin{tabular}[t]{l | c | c c c}
    \hline
     & ye- & \multicolumn{3}{|c}{$\ssttot$} \\
    Beam & ars & 0.01 & 0.05 & 0.001 \\
    \hline
    Fermi                & 2  & $0.104\pm 0.0060$                 & $0.501\pm 0.012\phantom{1}$                     & $0.012\pm 0.0034$  \\
    Fermi                & 8  & $0.104\pm 0.0032$                 & $0.501\pm 0.0061$                   & $0.012\pm 0.0021$  \\
    JHF $\nu$          & 2 & $(7.9\pm 0.37)\cdot 10^{-3}$ & $(3.2\pm 0.052)\cdot 10^{-2}$ & $(1.5\pm 0.32)\cdot 10^{-3}$  \\
    JHF $\nu$          & 8 & $(7.9\pm 0.25)\cdot 10^{-3}$ & $(3.2\pm 0.031)\cdot 10^{-2}$ & $(1.5\pm 0.24)\cdot 10^{-3}$  \\
    JHF $\bar{\nu}$ & 6 & $(2.8\pm 0.33)\cdot 10^{-3}$ & $(1.8\pm 0.044)\cdot 10^{-2}$ & $(0.0\pm 0.31)\cdot 10^{-3}$  \\
    \hline
  \end{tabular}
  \caption{Expected oscillation probability measurements for the various beams and input values.}
  \label{tab:measurements}
\end{table}

\begin{figure}[!p]
  \setlength{\abovecaptionskip}{0pt}
  \parbox[c]{\textwidth}{\centering{
      \framebox[1.05\width][c]
      {Input values:  $\ssttot$=0.01, $\delta = \pi / 2$.
      Contours are 1, 2, 3$\sigma$}}}
  \vskip 1ex
  \begin{minipage}[t]{0.5\textwidth}
    \flushright
    \includegraphics[width=\textwidth]{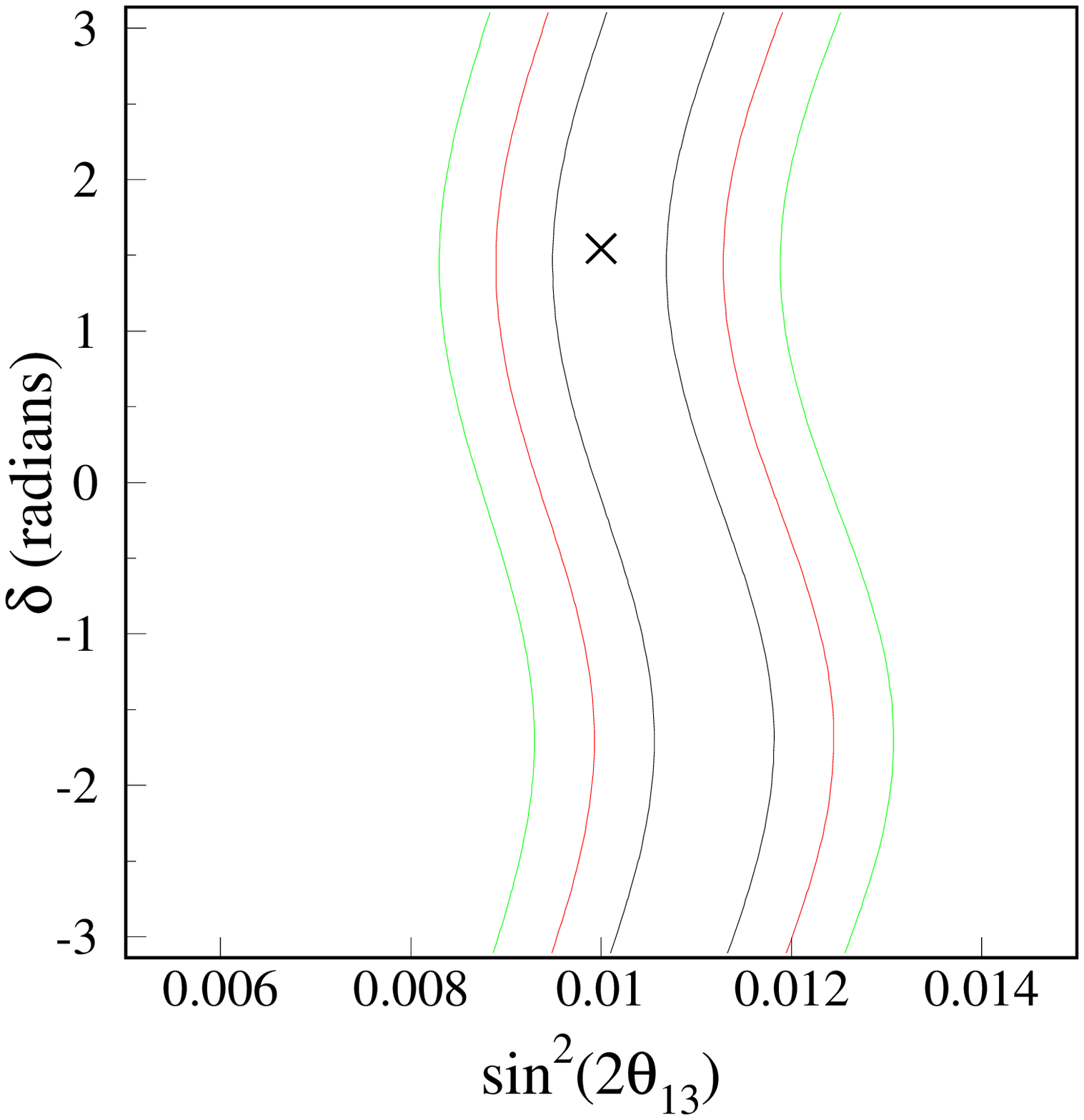}
    \parbox[t]{0.9\textwidth}{
      \caption{After 2 years with
        Fermilab beam.}
      \label{fig:tvd_01_f2}}
  \end{minipage}
  \begin{minipage}[t]{0.5\textwidth}
    \flushright
    \includegraphics[width=\textwidth]{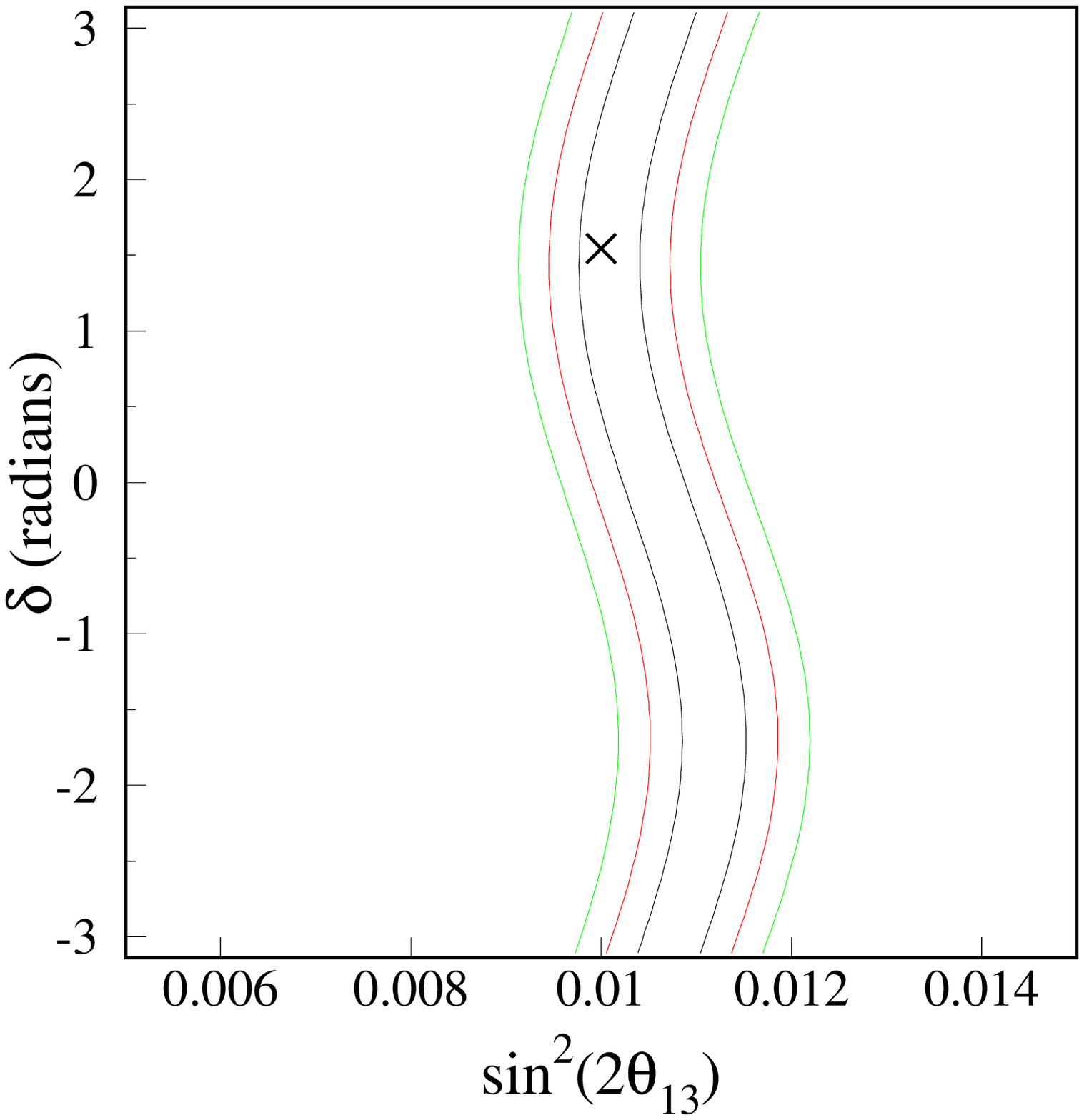}
    \parbox[t]{0.9\textwidth}{
      \caption{After 8 years with
        Fermilab beam.}
      \label{fig:tvd_01_f8}}
  \end{minipage}
  \vskip 2em
  \begin{minipage}[t]{0.5\textwidth}
    \flushright
    \includegraphics[width=\textwidth]{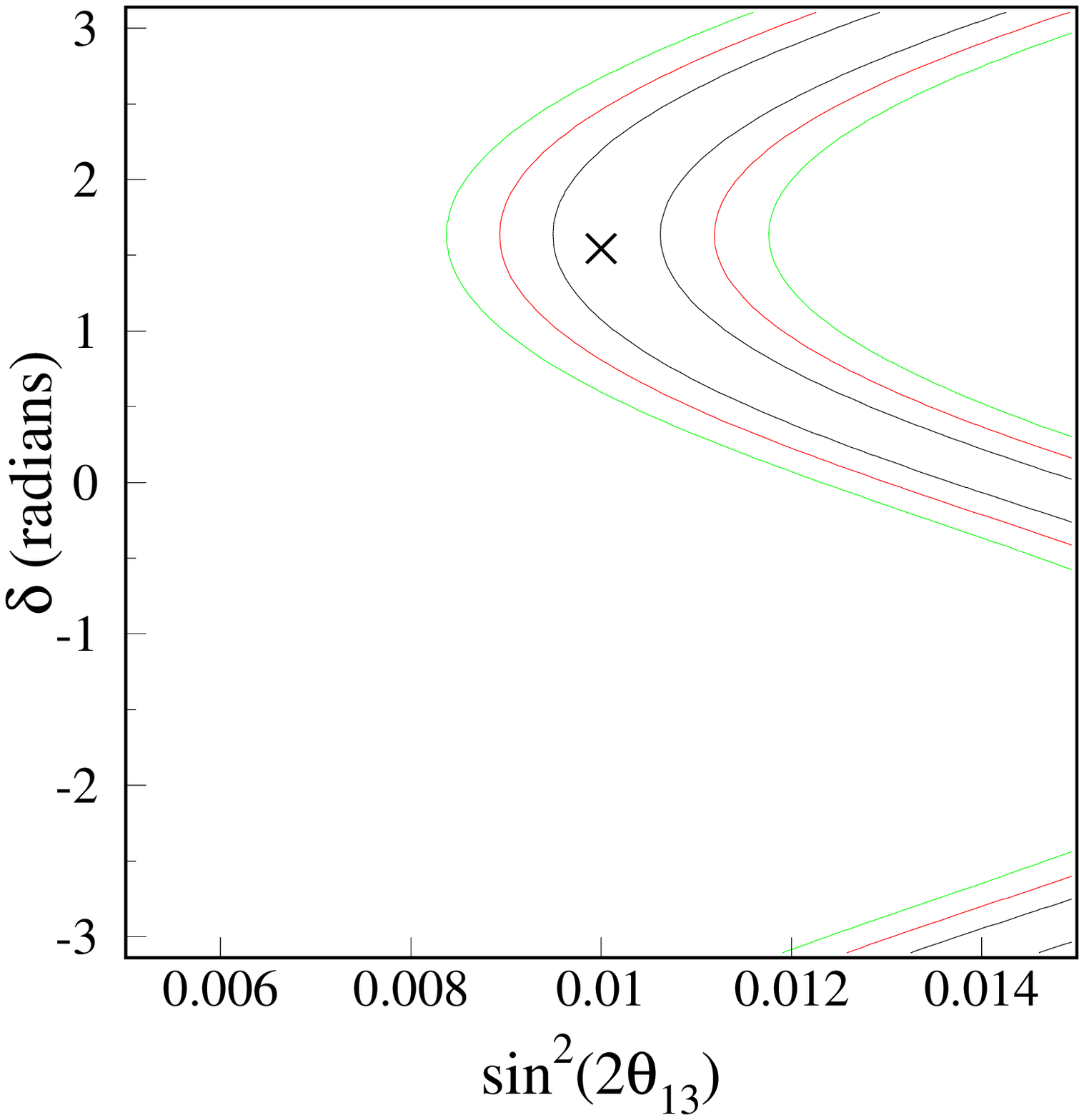}
    \parbox[t]{0.9\textwidth}{
      \caption{After 2 years with
        JHF beam.}
      \label{fig:tvd_01_k2}}
  \end{minipage}
  \begin{minipage}[t]{0.5\textwidth}
    \flushright
    \includegraphics[width=\textwidth]{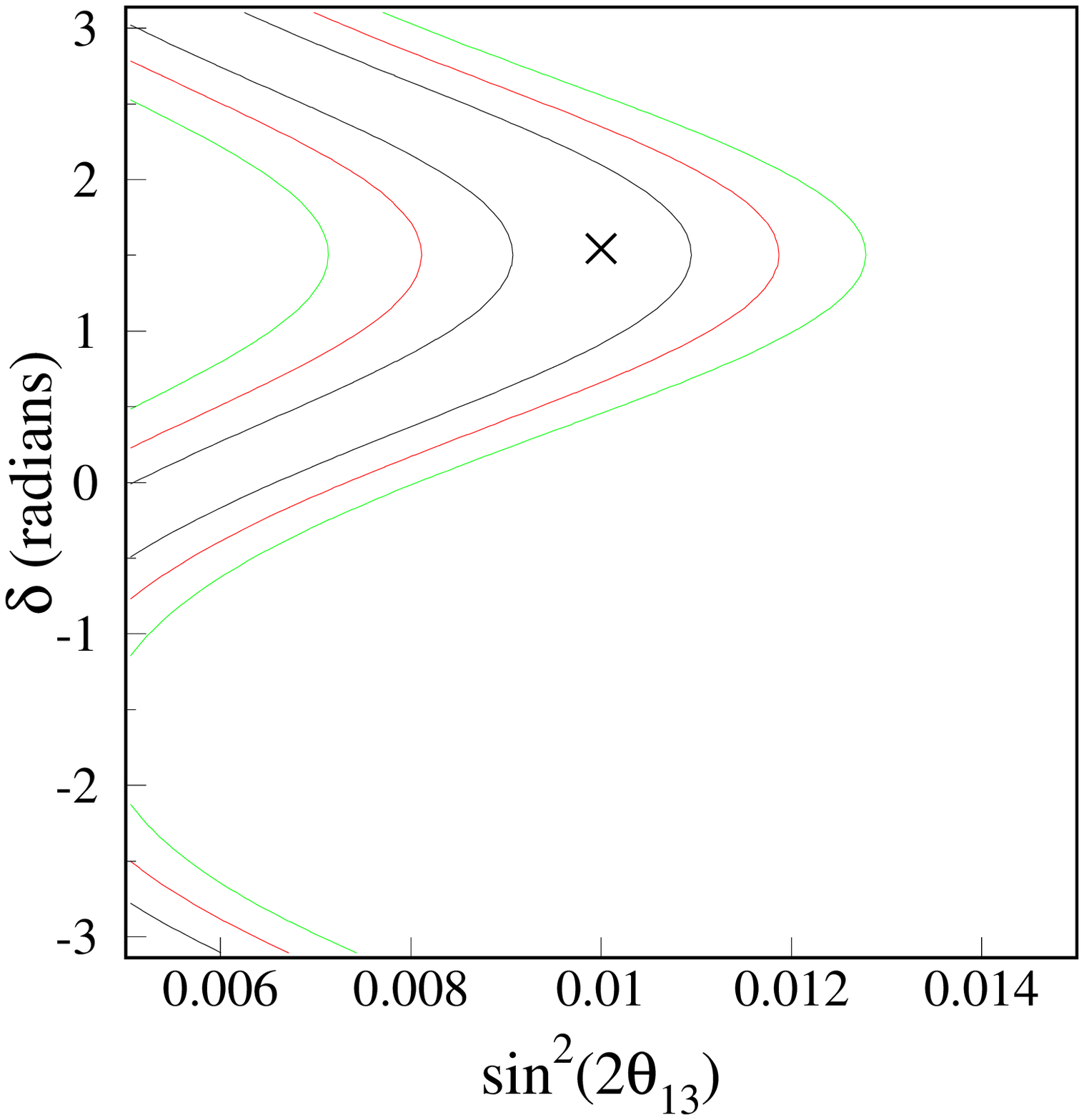}
    \parbox[t]{0.9\textwidth}{
      \caption{After 6 years with
        JHF antineutrino beam.}
    \label{fig:tvd_01_a2}}
  \end{minipage}
\end{figure}

\begin{figure}[!p]
  \setlength{\abovecaptionskip}{0pt}
  \parbox[c]{\textwidth}{\centering{
      \framebox[1.05\width][c]
      {Input values:  $\ssttot$=0.01, $\delta = \pi / 2$.
      Contours are 1, 2, 3$\sigma$}}}
  \vskip 1ex
  \begin{minipage}[t]{0.5\textwidth}
    \flushright
    \includegraphics[width=\textwidth]{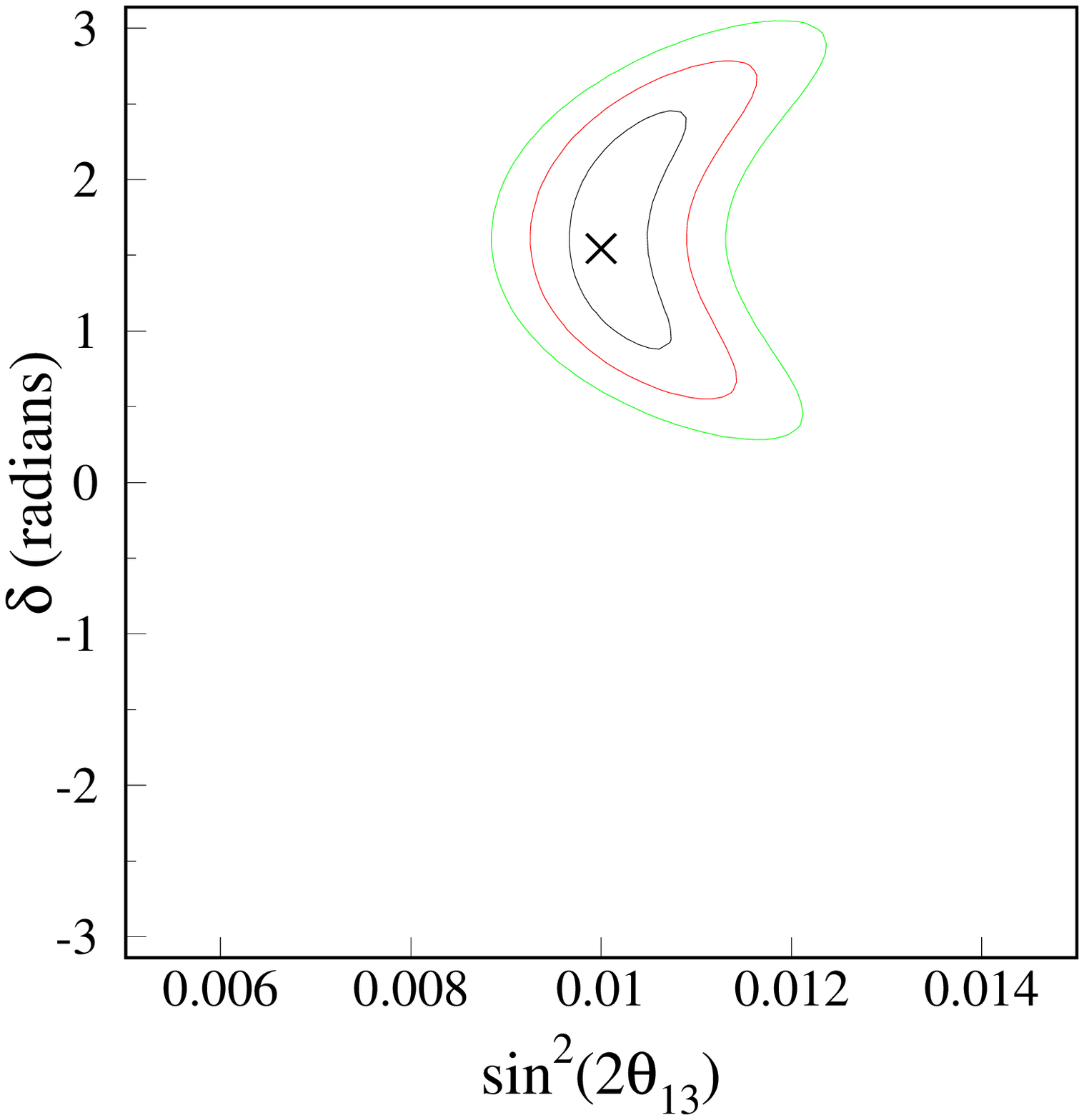}
    \parbox[t]{0.9\textwidth}{
      \caption{After 2 years with
        Fermilab and JHF beams.}
      \label{fig:tvd_01_f2_k2}}
  \end{minipage}
  \begin{minipage}[t]{0.5\textwidth}
    \flushright
    \includegraphics[width=\textwidth]{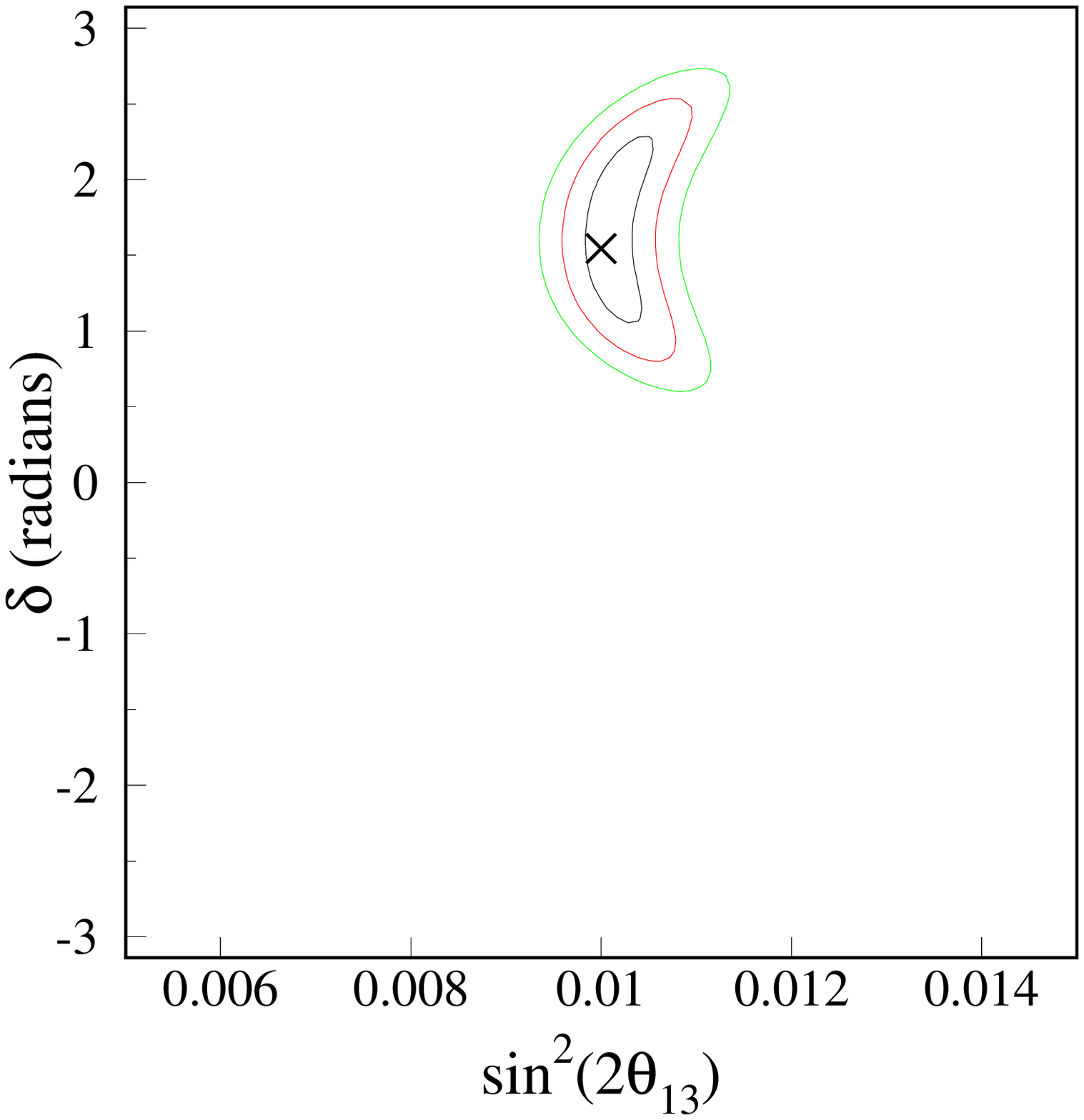}
    \parbox[t]{0.9\textwidth}{
      \caption{After 8 years with
        Fermilab and JHF beams.}
      \label{fig:tvd_01_f8_k8}}
  \end{minipage}
  \vskip 2em
  \begin{minipage}[t]{0.5\textwidth}
    \flushright
    \includegraphics[width=\textwidth]{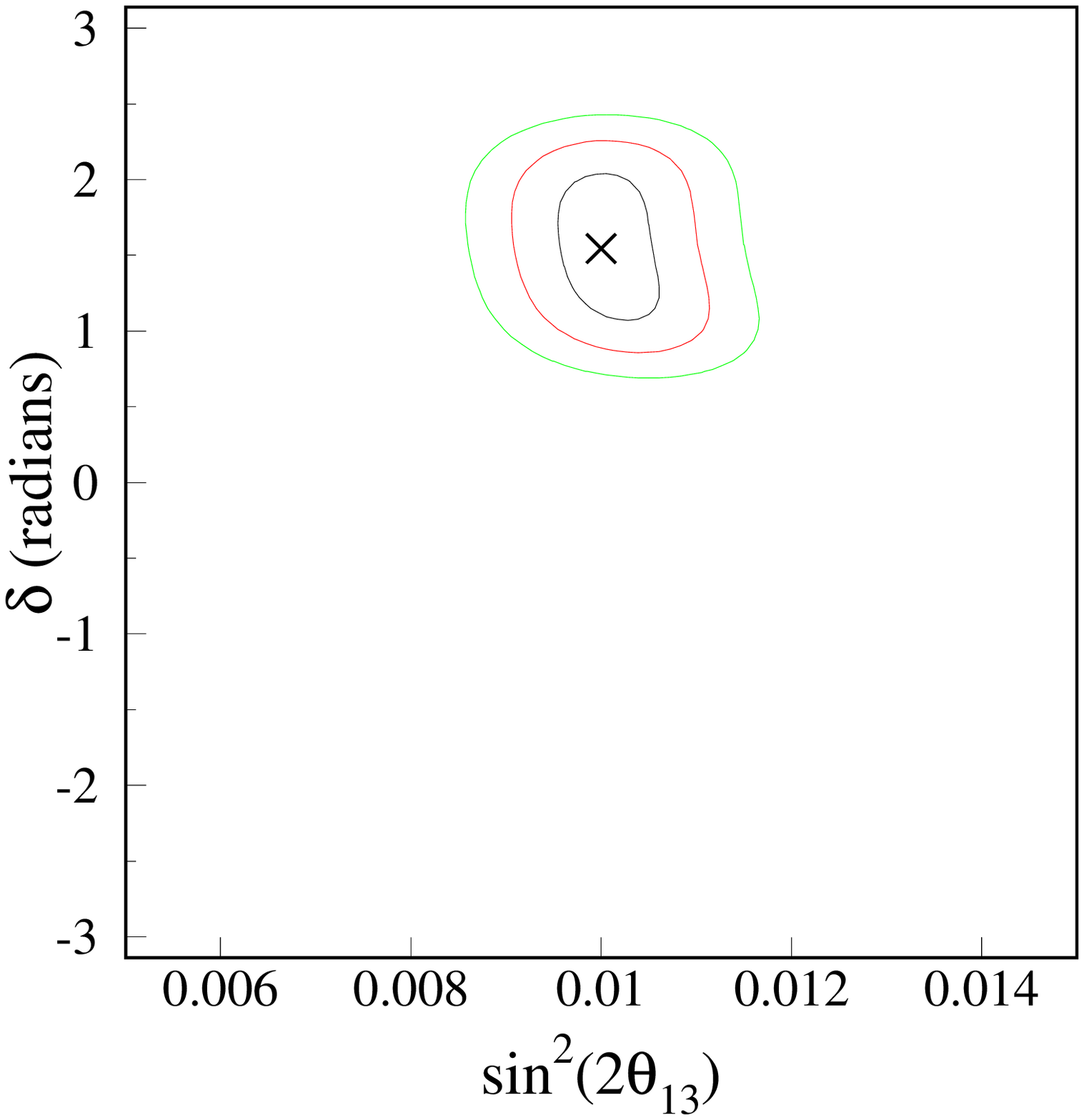}
    \parbox[t]{0.9\textwidth}{
      \caption{After 2 years with JHF beam and 6 years with JHF
        antineutrino beam.}
      \label{fig:tvd_01_k2_a2}}
  \end{minipage}
  \begin{minipage}[t]{0.5\textwidth}
    \flushright
    \includegraphics[width=\textwidth]{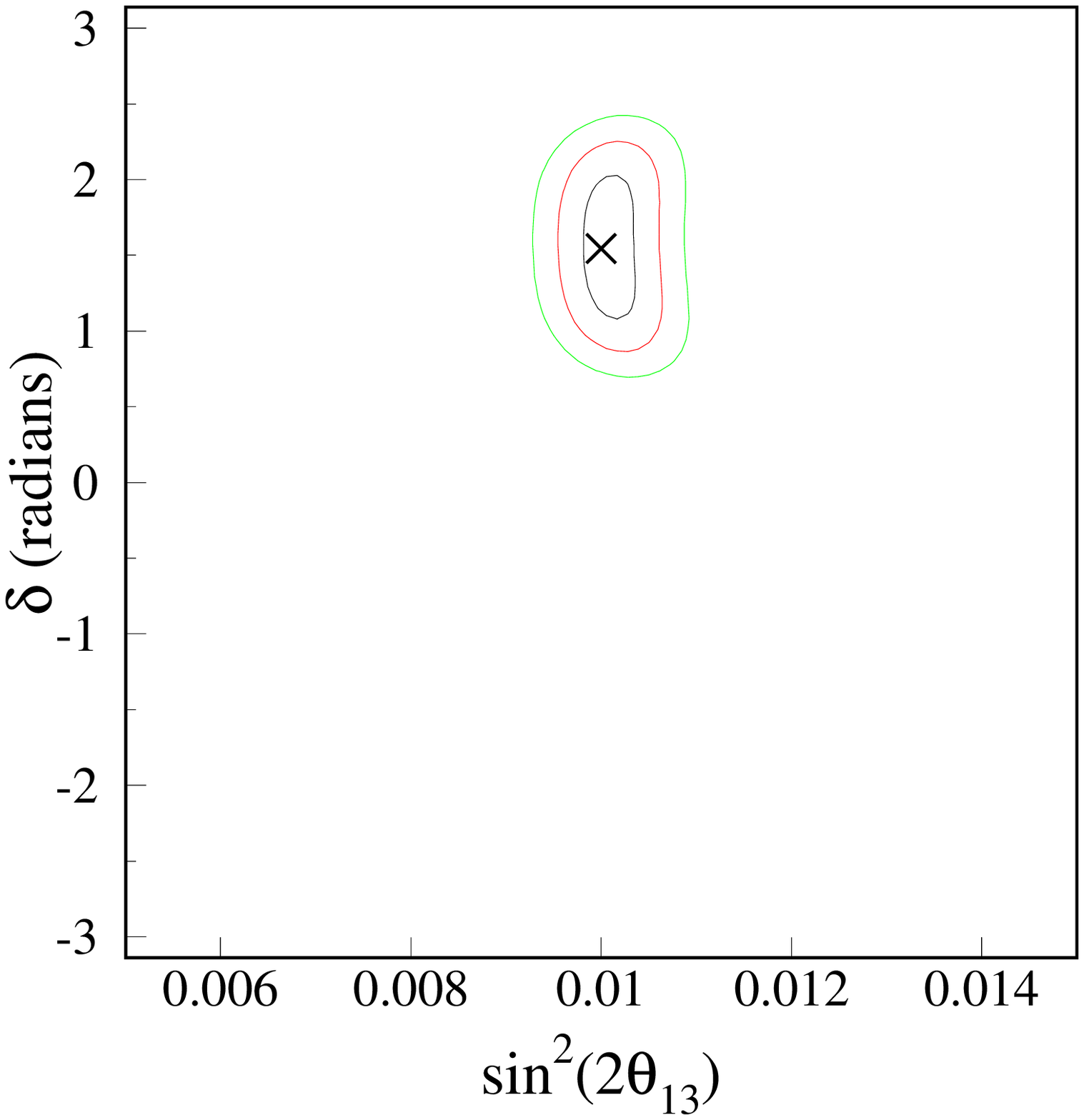}
    \parbox[t]{0.9\textwidth}{
      \caption{ After 8 years with
        Fermilab, 2 years with JHF, and 6 years with
        antineutrino beams.}
    \label{fig:tvd_01_f8_k2_a2}}
  \end{minipage}
\end{figure}

\begin{figure}[!p]
  \setlength{\abovecaptionskip}{0pt}
  \parbox[c]{\textwidth}{\centering{
      \framebox[1.05\width][c]
      {Input values:  $\ssttot$=0.05, $\delta = \pi / 2$.
      Contours are 1, 2, 3$\sigma$}}}
  \vskip 1ex
  \begin{minipage}[t]{0.5\textwidth}
    \flushright
    \includegraphics[width=\textwidth]{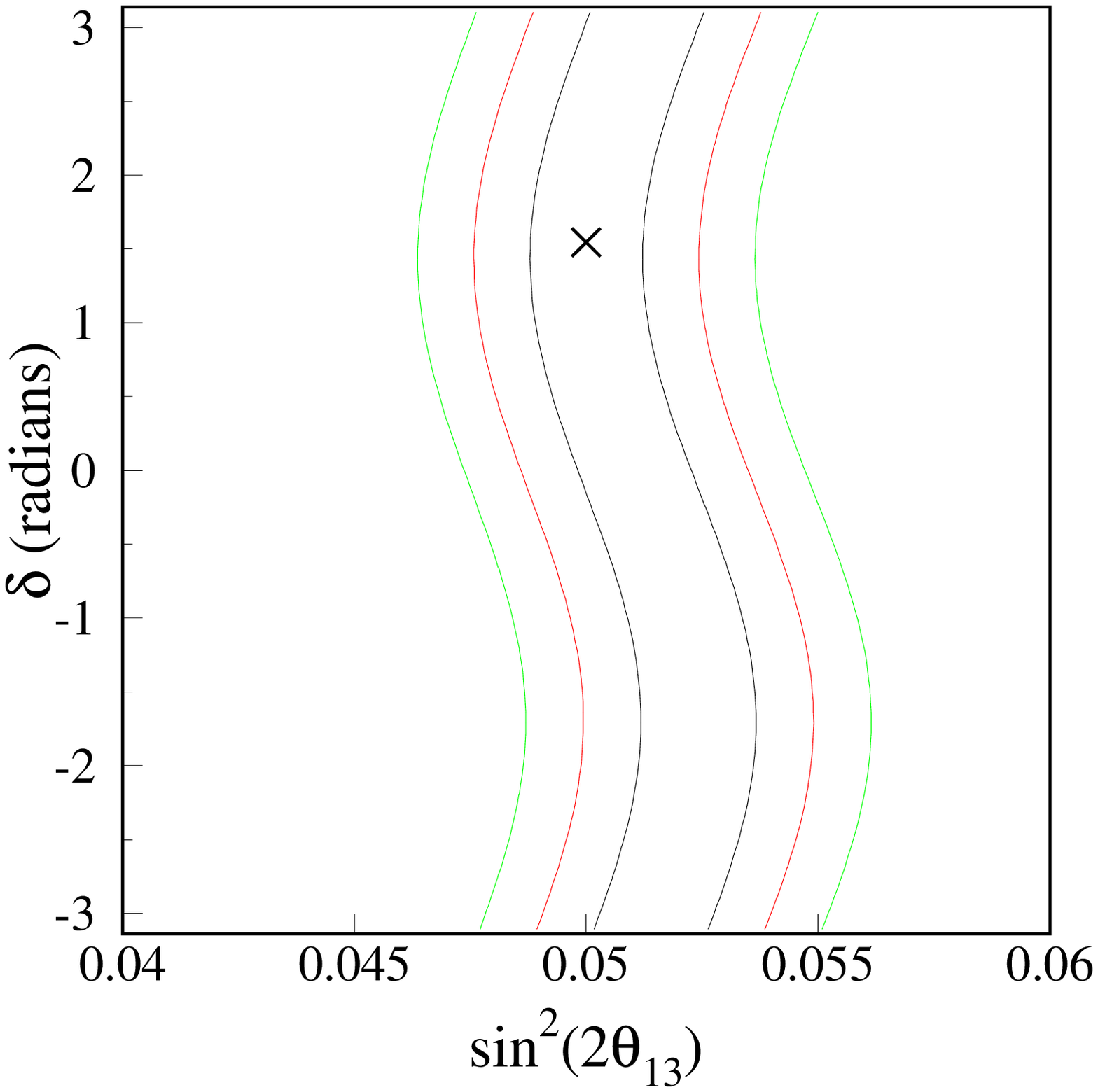}
    \parbox[t]{0.9\textwidth}{
      \caption{After 2 years with
        Fermilab beam.}
      \label{fig:tvd_05_f2}}
  \end{minipage}
  \begin{minipage}[t]{0.5\textwidth}
    \flushright
    \includegraphics[width=\textwidth]{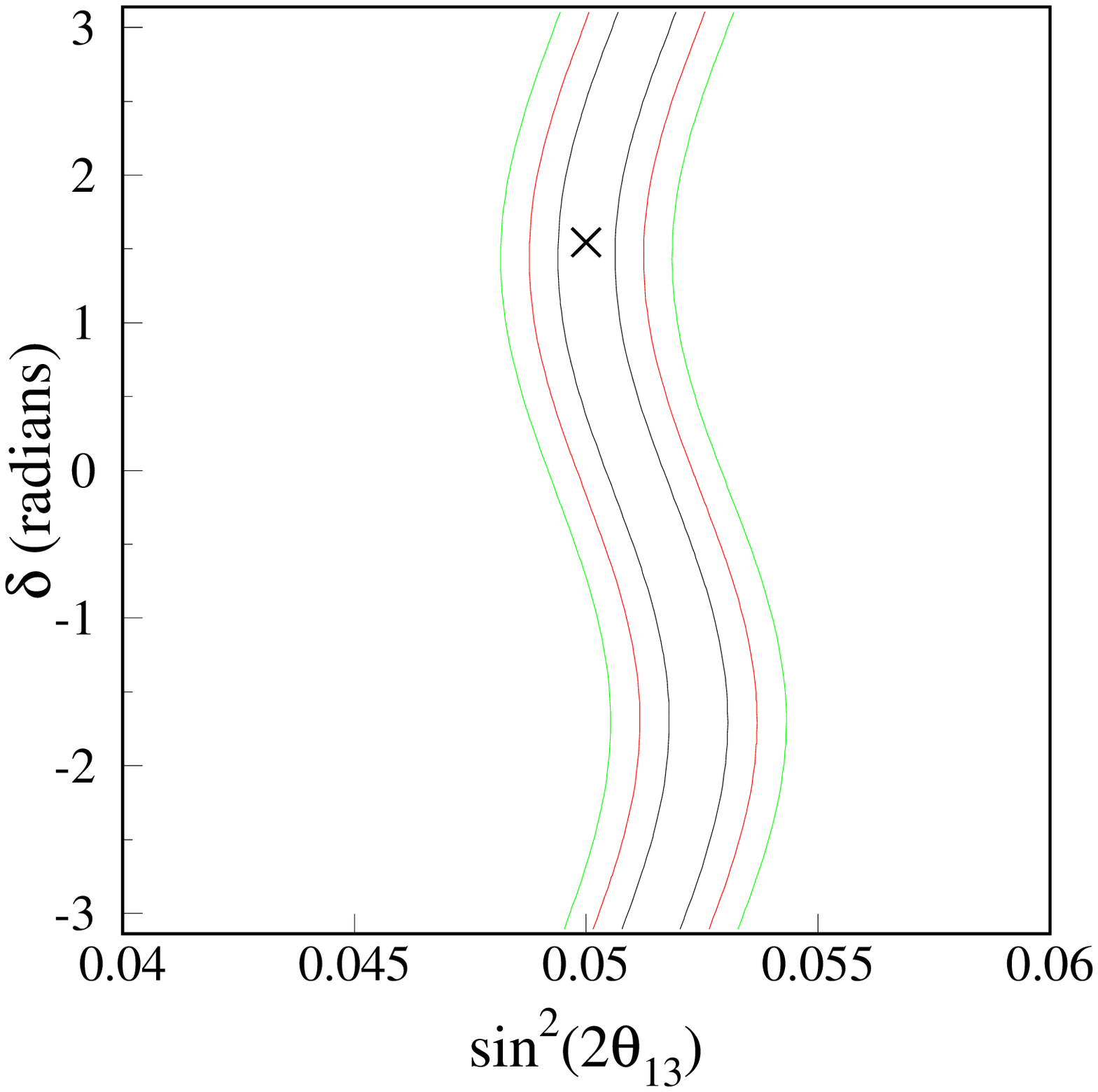}
    \parbox[t]{0.9\textwidth}{
      \caption{After 8 years with
        Fermilab beam.}
      \label{fig:tvd_05_f8}}
  \end{minipage}
  \vskip 2em
  \begin{minipage}[t]{0.5\textwidth}
    \flushright
    \includegraphics[width=\textwidth]{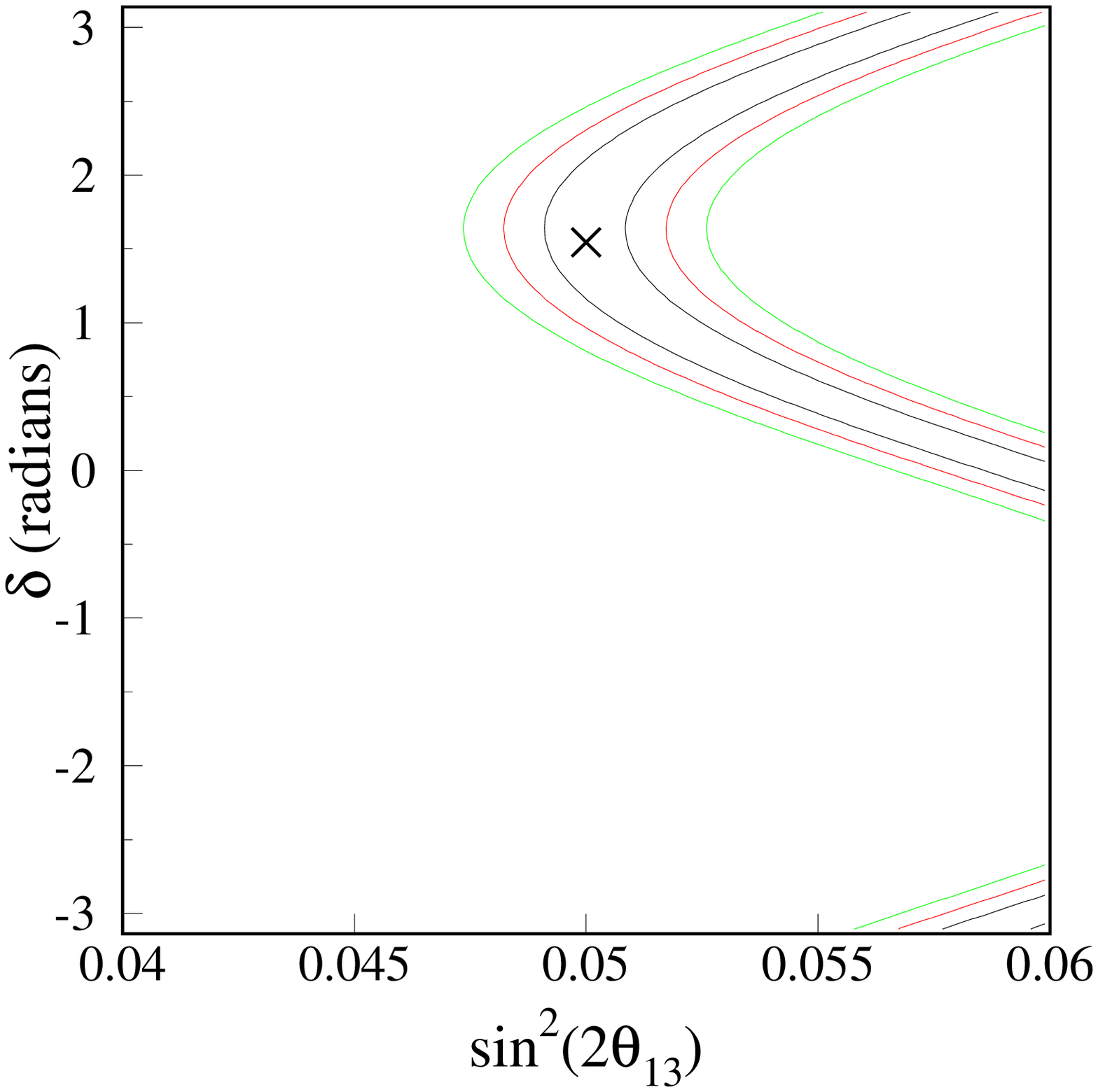}
    \parbox[t]{0.9\textwidth}{
      \caption{After 2 years with
        JHF beam.}
      \label{fig:tvd_05_k2}}
  \end{minipage}
  \begin{minipage}[t]{0.5\textwidth}
    \flushright
    \includegraphics[width=\textwidth]{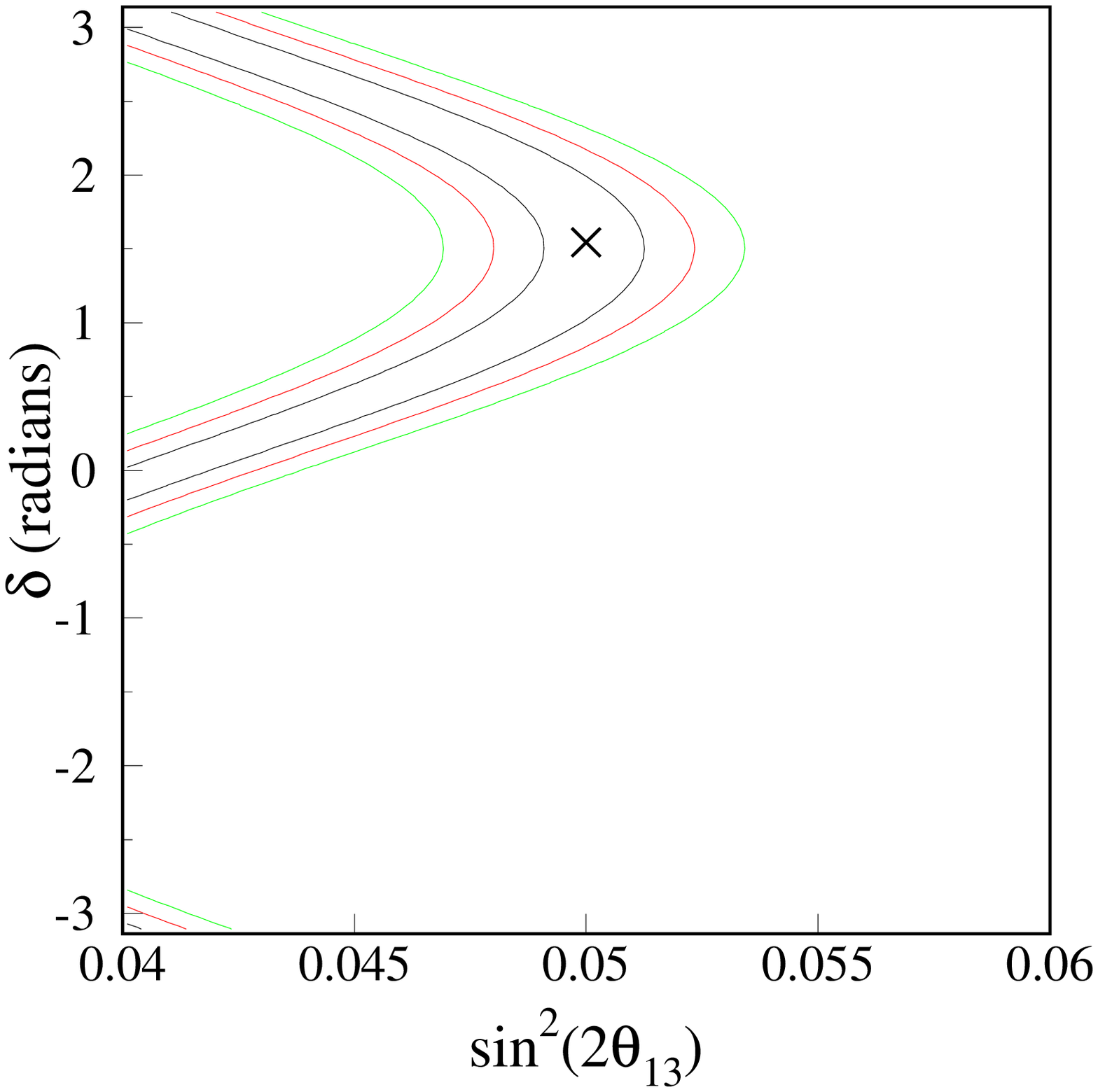}
    \parbox[t]{0.9\textwidth}{
      \caption{After 6 years with
        JHF antineutrino beam.}
    \label{fig:tvd_05_a2}}
  \end{minipage}
\end{figure}

\begin{figure}[!p]
  \setlength{\abovecaptionskip}{0pt}
  \parbox[c]{\textwidth}{\centering{
      \framebox[1.05\width][c]
      {Input values:  $\ssttot$=0.05, $\delta = \pi / 2$.
      Contours are 1, 2, 3$\sigma$}}}
  \vskip 1ex
  \begin{minipage}[t]{0.5\textwidth}
    \flushright
    \includegraphics[width=\textwidth]{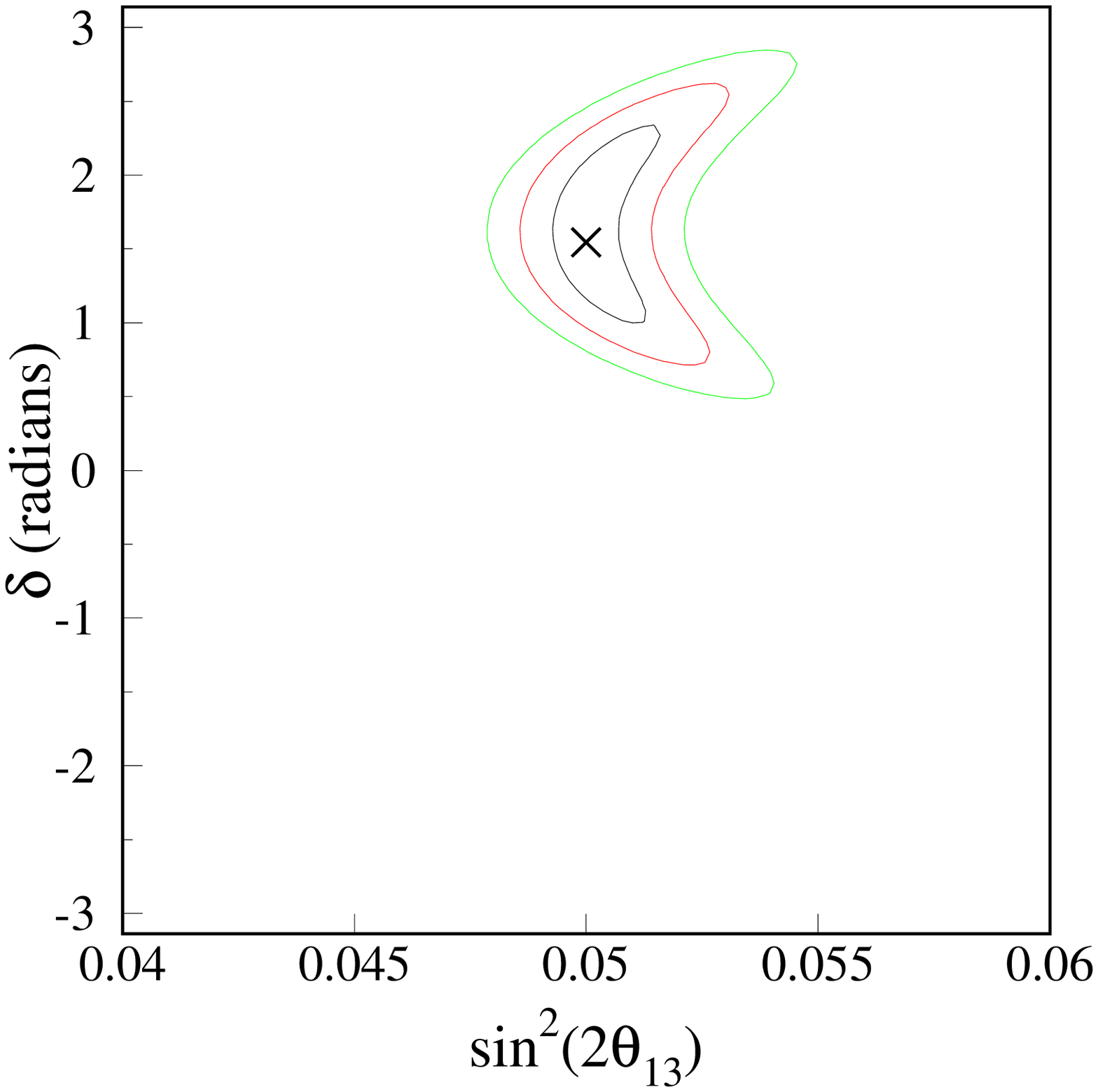}
    \parbox[t]{0.9\textwidth}{
      \caption{After 2 years with
        Fermilab and JHF beams.}
      \label{fig:tvd_05_f2_k2}}
  \end{minipage}
  \begin{minipage}[t]{0.5\textwidth}
    \flushright
    \includegraphics[width=\textwidth]{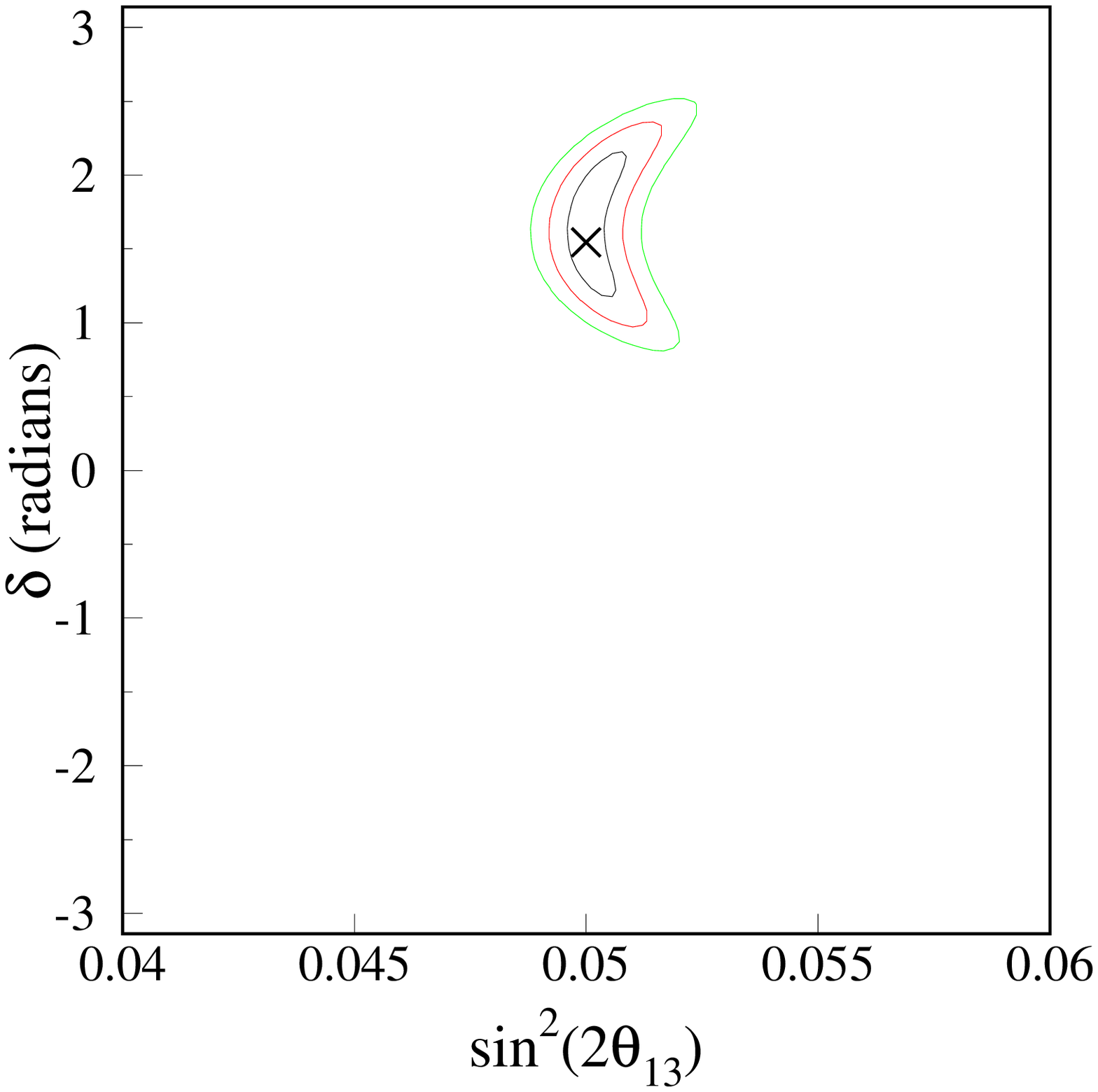}
    \parbox[t]{0.9\textwidth}{
      \caption{After 8 years with
        Fermilab and JHF beams.}
      \label{fig:tvd_05_f8_k8}}
  \end{minipage}
  \vskip 2em
  \begin{minipage}[t]{0.5\textwidth}
    \flushright
    \includegraphics[width=\textwidth]{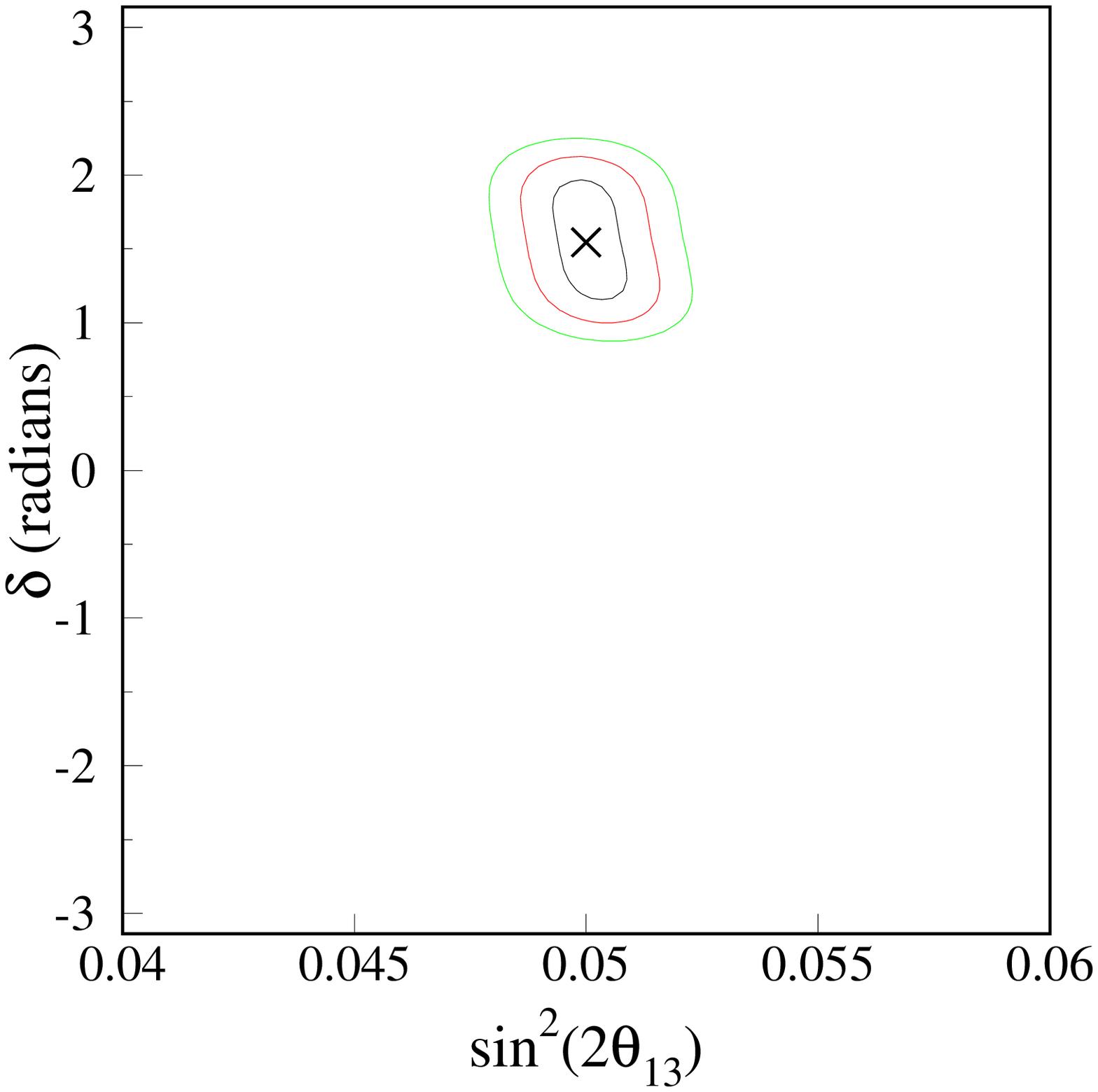}
    \parbox[t]{0.9\textwidth}{
      \caption{After 2 years with JHF beam and 6 years with JHF
        antineutrino beam.}
      \label{fig:tvd_05_k2_a2}}
  \end{minipage}
  \begin{minipage}[t]{0.5\textwidth}
    \flushright
    \includegraphics[width=\textwidth]{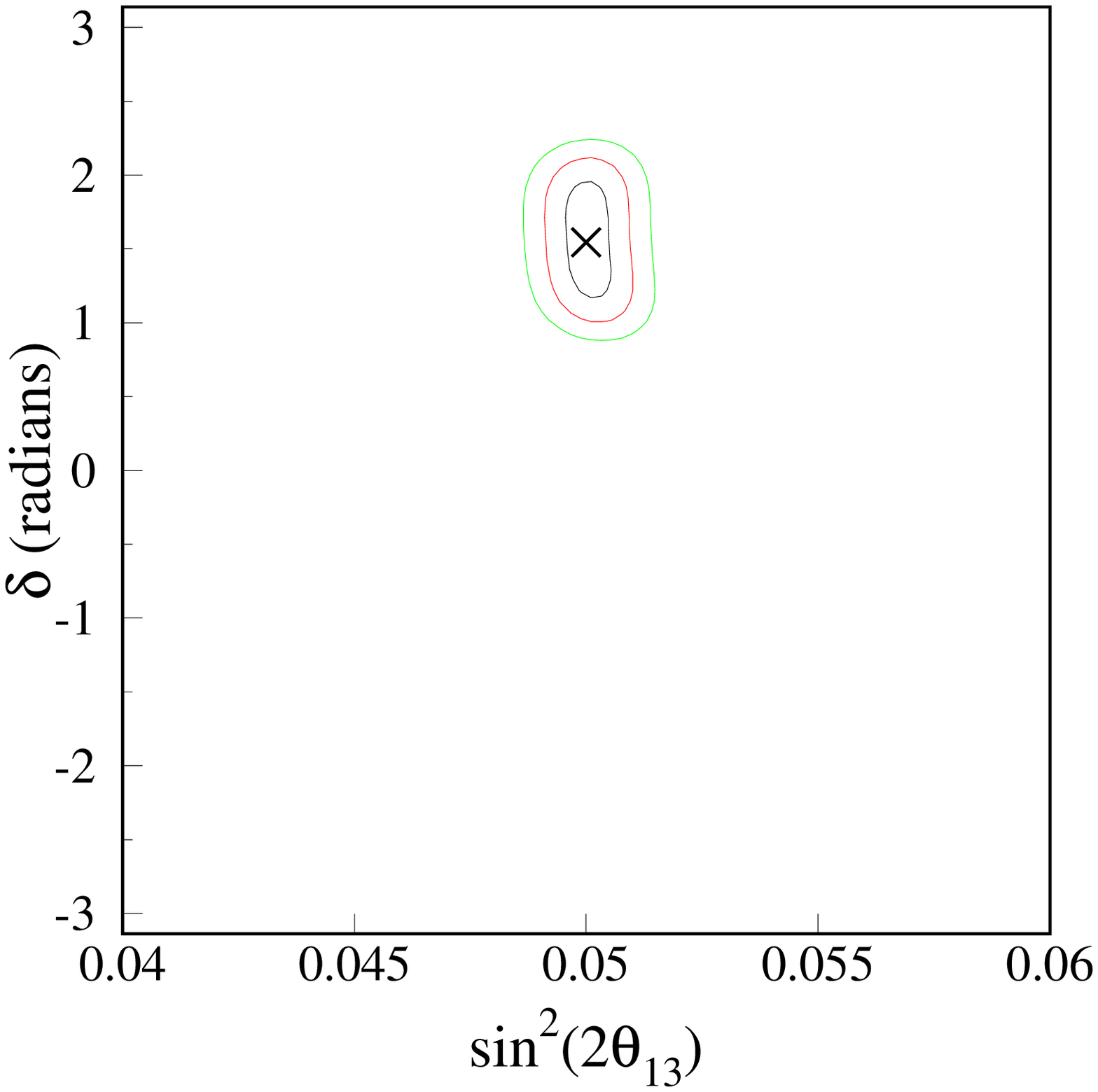}
    \parbox[t]{0.9\textwidth}{
      \caption{ After 8 years with
        Fermilab, 2 years with JHF, and 6 years with
        antineutrino beams.}
    \label{fig:tvd_05_f8_k2_a2}}
  \end{minipage}
\end{figure}

\begin{figure}[!p]
  \setlength{\abovecaptionskip}{0pt}
  \parbox[c]{\textwidth}{\centering{
      \framebox[1.05\width][c]
      {Input values:  $\ssttot$=0.001, $\delta = \pi / 2$.
      Contours are 1, 2, 3$\sigma$}}}
  \vskip 1ex
  \begin{minipage}[t]{0.5\textwidth}
    \flushright
    \includegraphics[width=\textwidth]{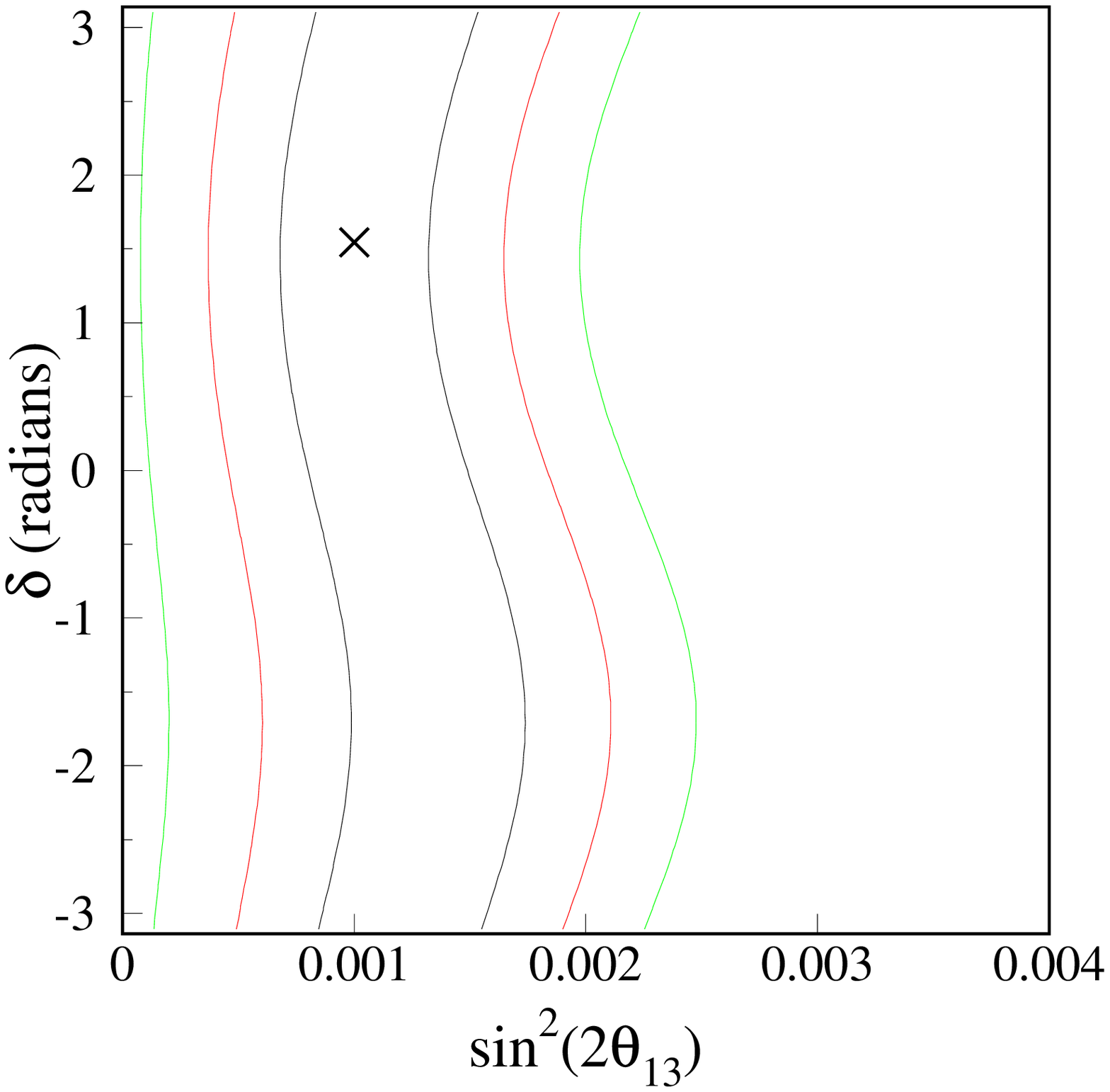}
    \parbox[t]{0.9\textwidth}{
      \caption{After 2 years with
        Fermilab beam.}
      \label{fig:tvd_001_f2}}
  \end{minipage}
  \begin{minipage}[t]{0.5\textwidth}
    \flushright
    \includegraphics[width=\textwidth]{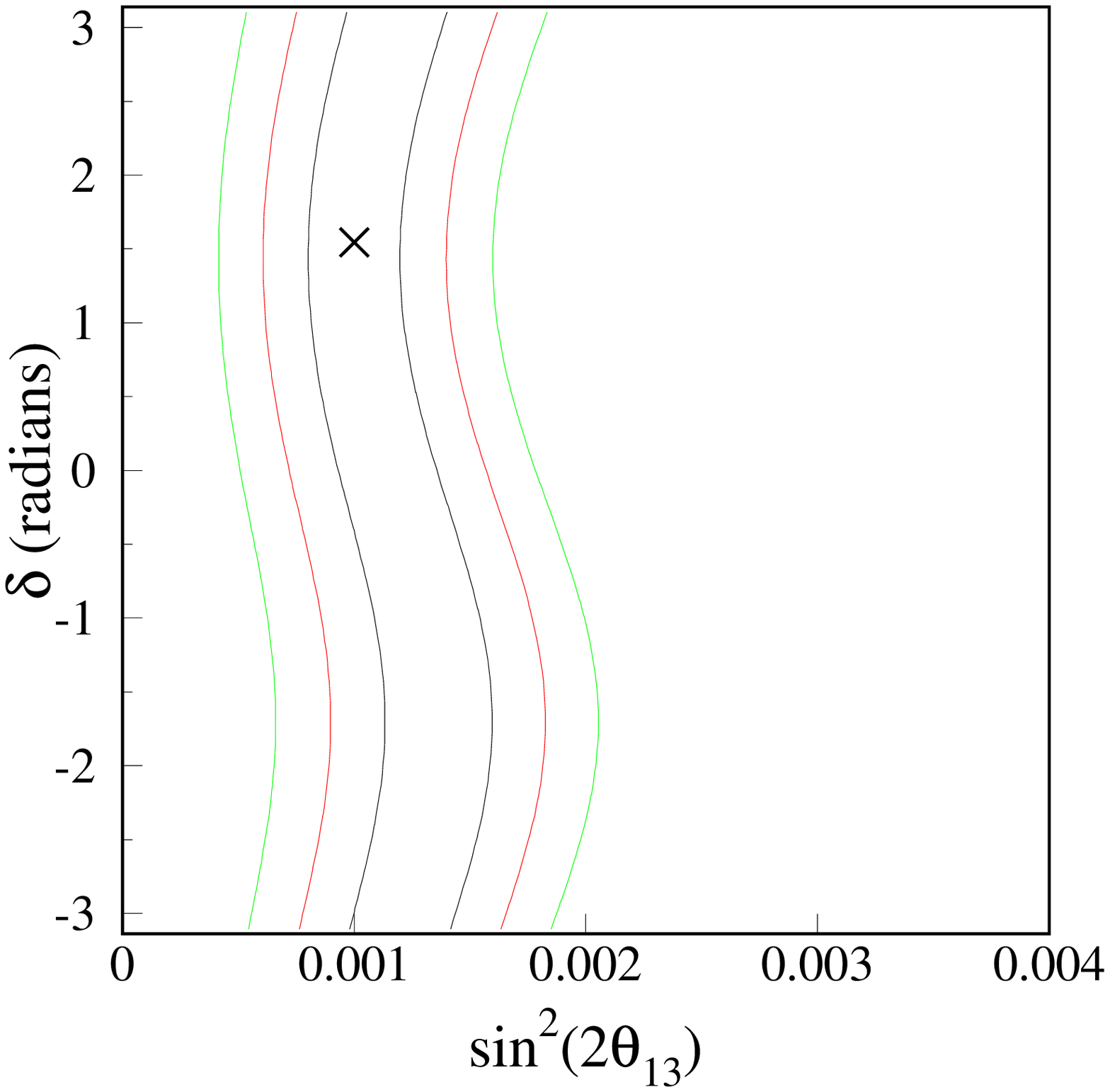}
    \parbox[t]{0.9\textwidth}{
      \caption{After 8 years with
        Fermilab beam.}
      \label{fig:tvd_001_f8}}
  \end{minipage}
  \vskip 2em
  \begin{minipage}[t]{0.5\textwidth}
    \flushright
    \includegraphics[width=\textwidth]{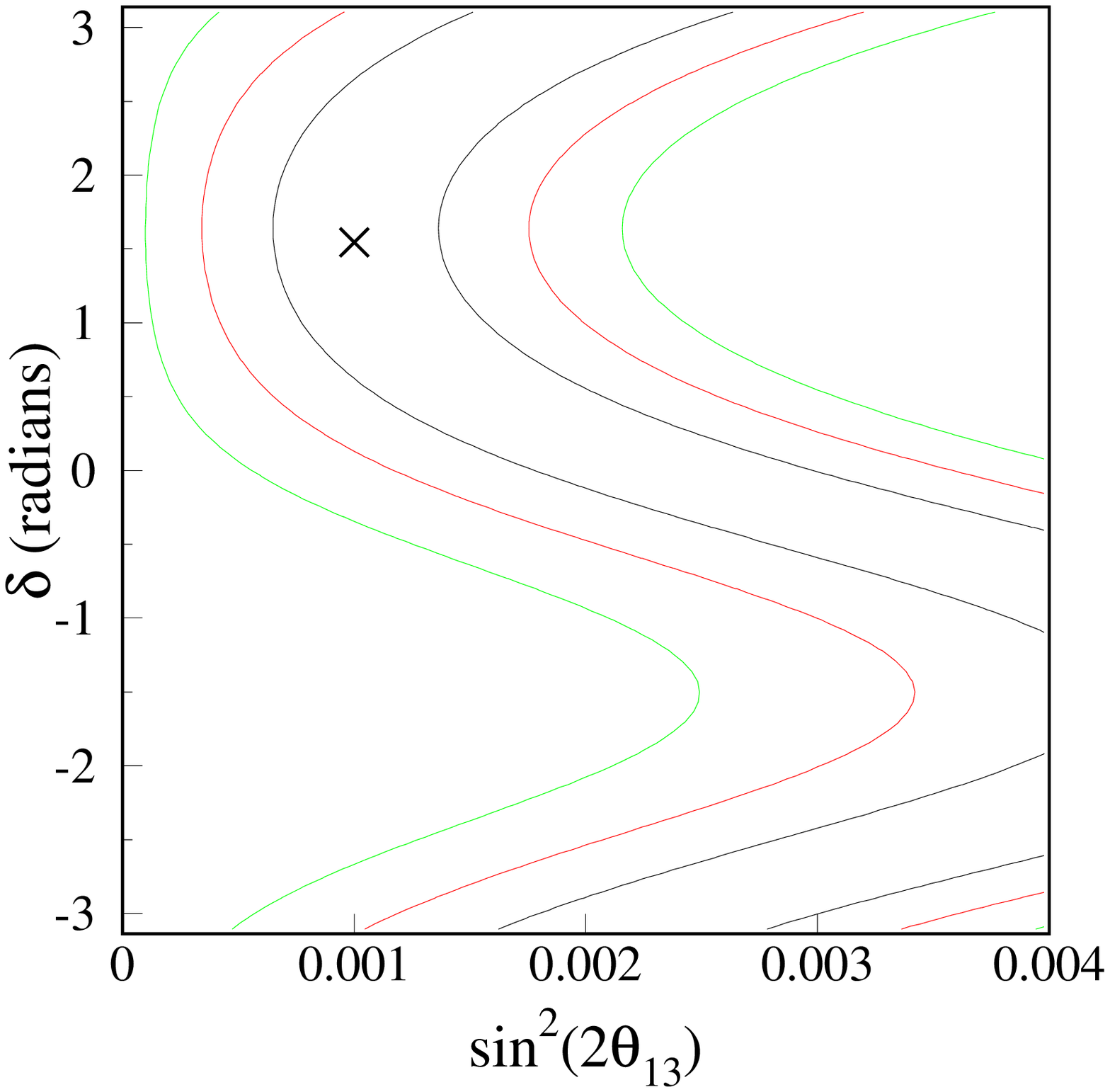}
    \parbox[t]{0.9\textwidth}{
      \caption{After 2 years with
        JHF beam.}
      \label{fig:tvd_001_k2}}
  \end{minipage}
  \begin{minipage}[t]{0.5\textwidth}
    \flushright
    \includegraphics[width=\textwidth]{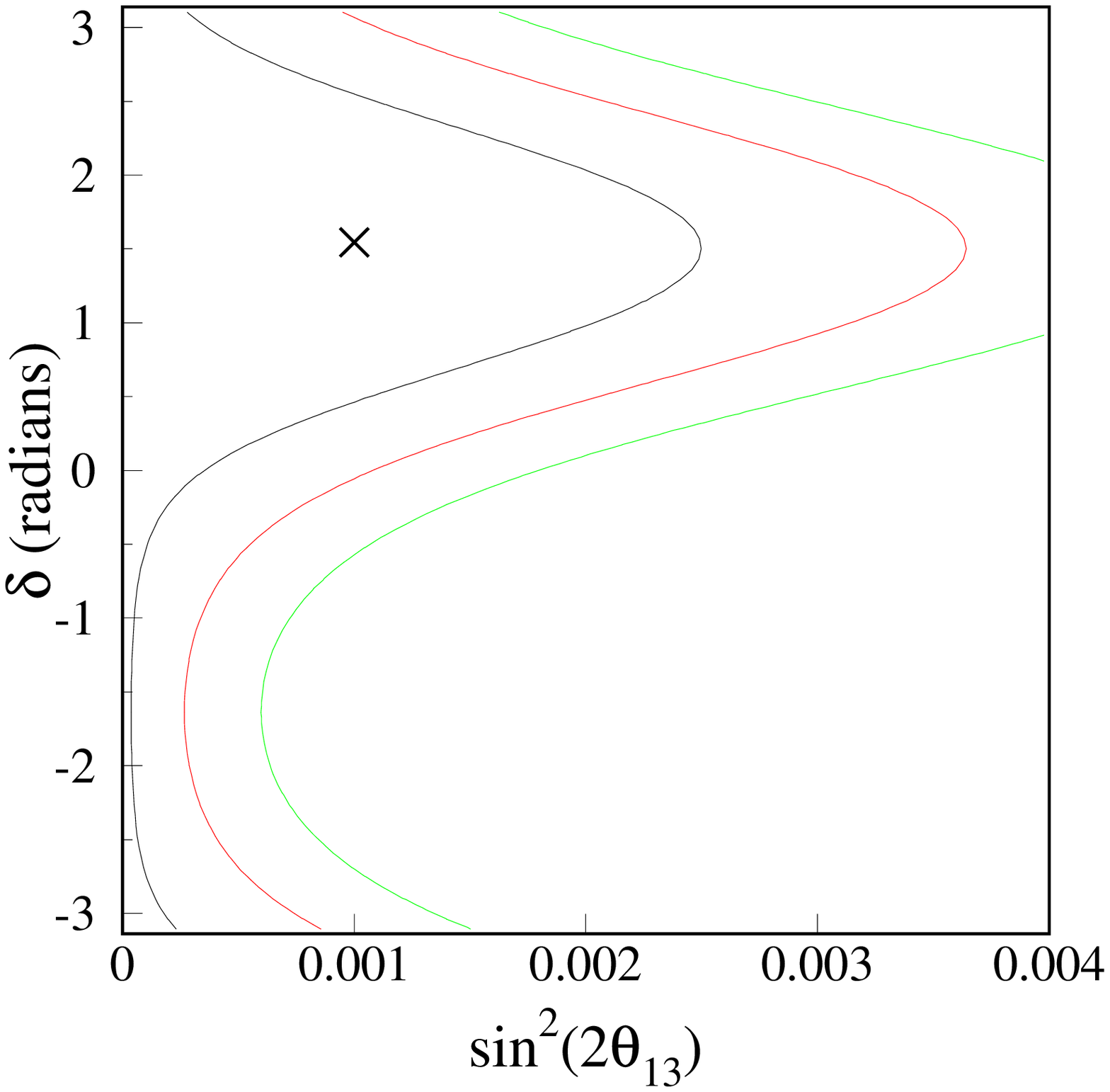}
    \parbox[t]{0.9\textwidth}{
      \caption{After 6 years with
        JHF antineutrino beam.}
    \label{fig:tvd_001_a2}}
  \end{minipage}
\end{figure}

\begin{figure}[!p]
  \setlength{\abovecaptionskip}{0pt}
  \parbox[c]{\textwidth}{\centering{
      \framebox[1.01\width][c]
      {Input values:  $\ssttot$=0.001, $\delta = \pi / 2$.
      Contours are 1, 2, 3$\sigma$}}}
  \vskip 1ex
  \begin{minipage}[t]{0.5\textwidth}
    \flushright
    \includegraphics[width=\textwidth]{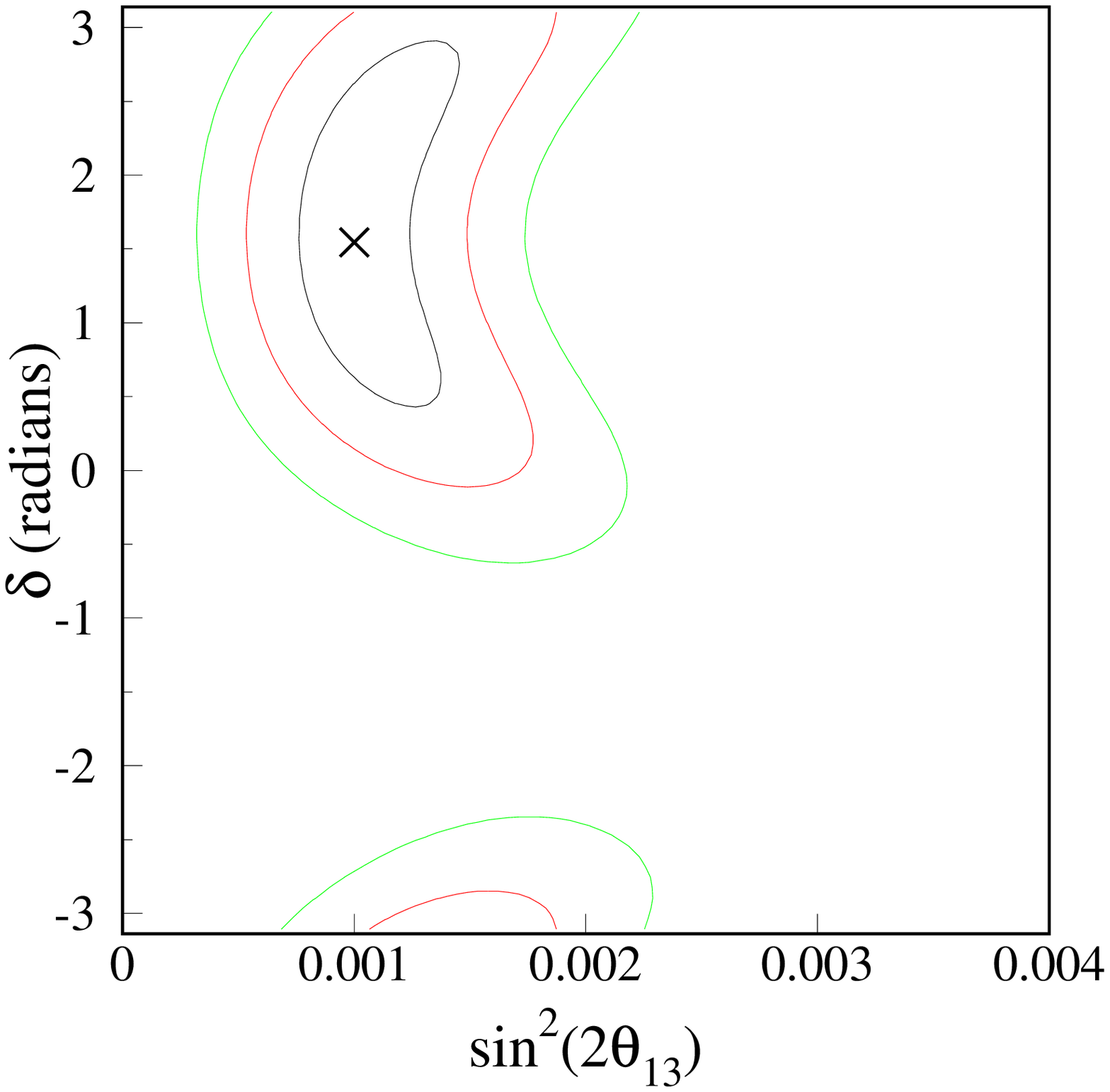}
    \parbox[t]{0.9\textwidth}{
      \caption{After 2 years with
        Fermilab and JHF beams.}
      \label{fig:tvd_001_f2_k2}}
  \end{minipage}
  \begin{minipage}[t]{0.5\textwidth}
    \flushright
    \includegraphics[width=\textwidth]{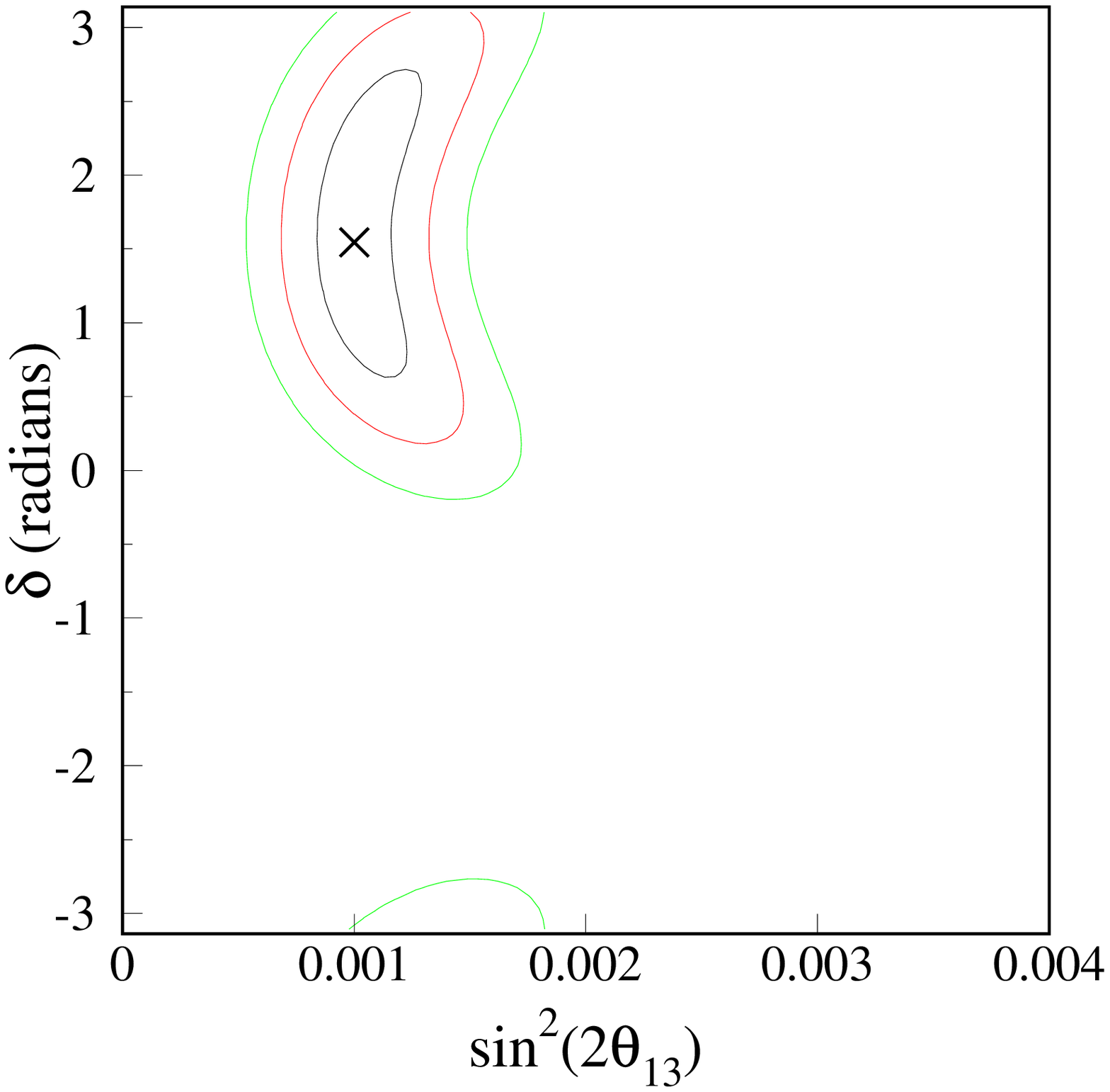}
    \parbox[t]{0.9\textwidth}{
      \caption{After 8 years with
        Fermilab and JHF beams.}
      \label{fig:tvd_001_f8_k8}}
  \end{minipage}
  \vskip 2em
  \begin{minipage}[t]{0.5\textwidth}
    \flushright
    \includegraphics[width=\textwidth]{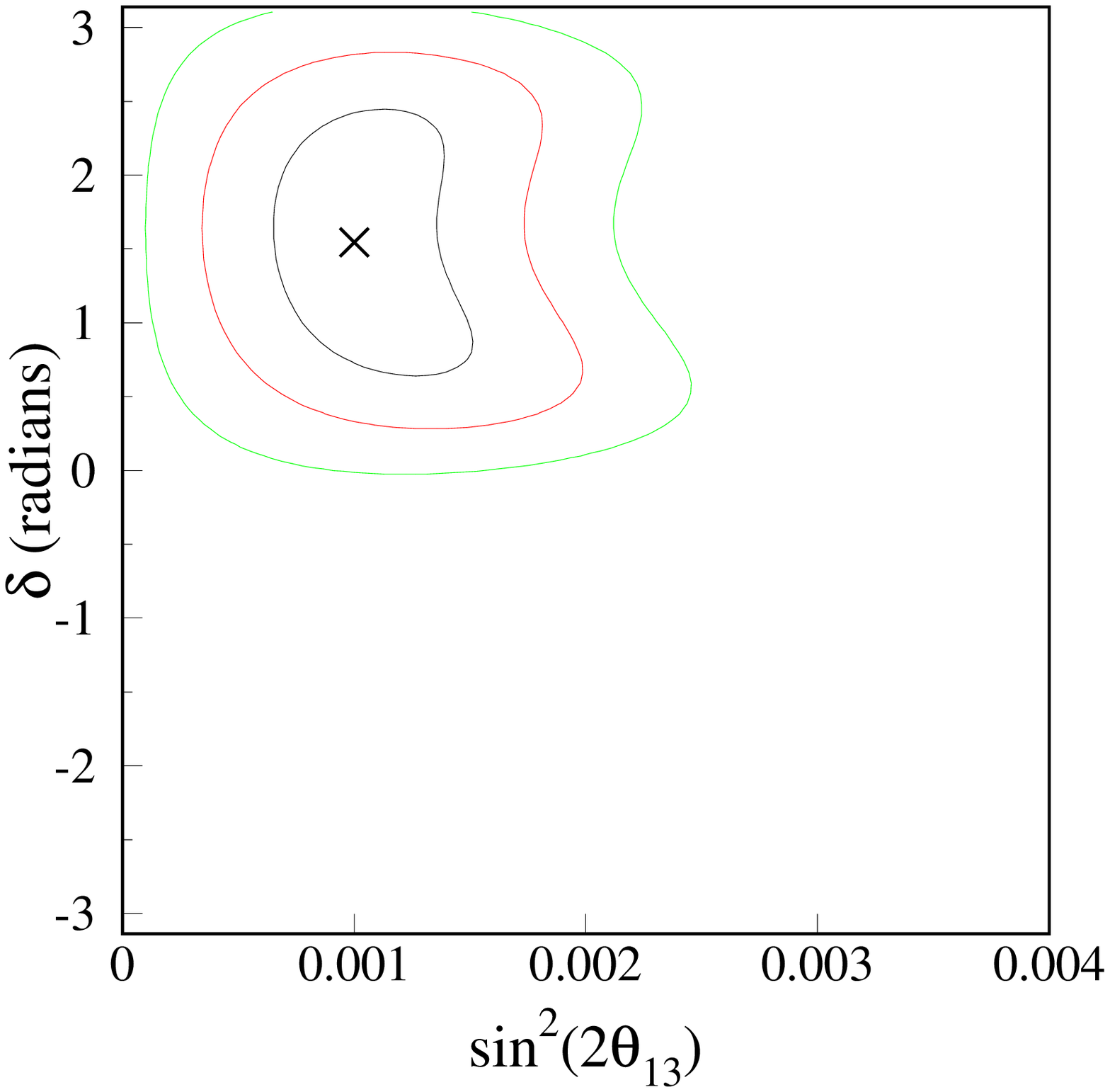}
    \parbox[t]{0.9\textwidth}{
      \caption{After 2 years with JHF beam and 6 years with JHF
        antineutrino beam.}
      \label{fig:tvd_001_k2_a2}}
  \end{minipage}
  \begin{minipage}[t]{0.5\textwidth}
    \flushright
    \includegraphics[width=\textwidth]{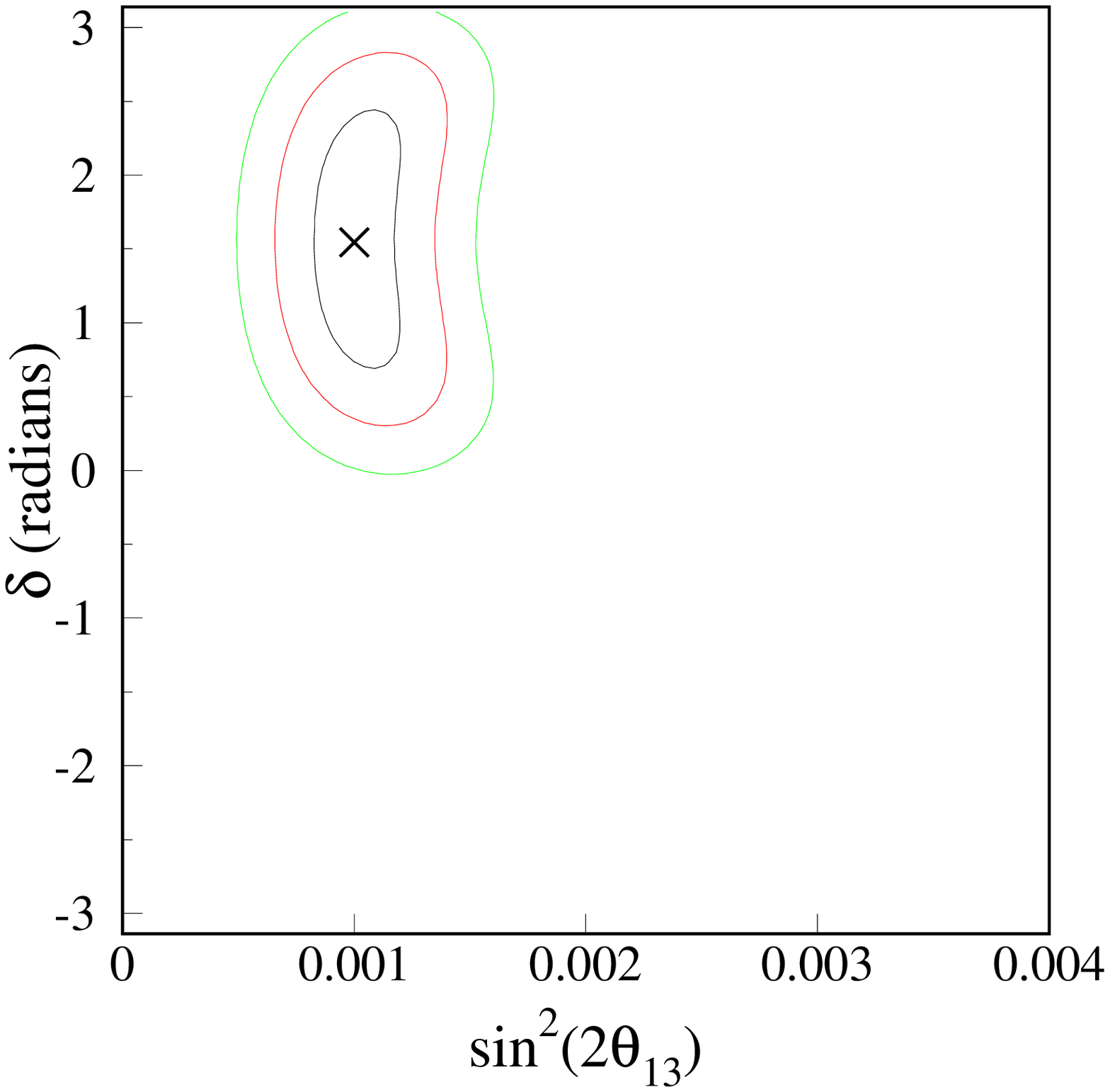}
    \parbox[t]{0.9\textwidth}{
      \caption{ After 8 years with
        Fermilab, 2 years with JHF, and 6 years with
        antineutrino beams.}
    \label{fig:tvd_001_f8_k2_a2}}
  \end{minipage}
\end{figure}

Constraints in the $\ssttot$-$\delta$ plane from various combinations
of these measurements are shown in Figs.~\ref{fig:tvd_01_f2}
through~~\ref{fig:tvd_001_f8_k2_a2}.  For each input value of
$\ssttot$, maximal CP violation can be detected at $3\sigma$ or
better, demonstrating a sensitivity over an impressive range of
parameter space.  This is possible since, as illustrated in
Fig.~\ref{fig:cp}, the maximum CP asymmetry is larger if $\ssttot$ is
smaller, so the larger asymmetry provides statistical compensation for
the smaller signal.  Over most of this range, 2 years with the
Fermilab and JHF beams combined is enough to reach the $3\sigma$
criterion.  The constraints are improved with an 8 year run during
which the JHF beam runs in antineutrino mode for 6 years.  If the JHF
antineutrino beam does not provide adequate performance, an 8 year run
with Fermilab and JHF neutrino beams provides constraints almost as
effective.

\section{Summary and conclusions}
\label{sec:conc}

We have investigated the physics potential of very long baseline
experiments designed to measure $\nu_\mu \to \nu_e$ oscillation
probabilities.  The principles of our design
are to tune the beam spectrum to the resonance energy for the matter
effect, and to have the spectrum cut off rapidly above this energy.
The matter effect amplifies the signal, and the cut-off suppresses backgrounds
which feed-down from higher energy.  The signal-to-noise ratio is
potentially better than for any other conventional $\nu_\mu$ beam
experiment.

The measurements from long- and short-baseline experiments are very
complementary.  The short-baseline experiment measures a CP-violating
combination of atmospheric and solar oscillations.  The relatively
small matter effects can still confuse the interpretation of the
measurements.  At long baselines, solar oscillations play a relatively
much smaller role.  The long-baseline experiment determines the sign
of $\Delta M^2$, constrains the matter effects, and constrains the
size of the atmospheric $\nu_\mu \to \nu_e$ oscillation alone.

In particular, we have investigated the capabilities of a neutrino
beam from Fermilab aimed at the Super-K detector in Japan, 9300 km
distant.  At this baseline, neutrinos at the resonance energy have an
oscillation probability 20 times the maximum in vacuum.  We have used
a simple model to estimate the signal efficiency and background, and
it will be important to check this with a full simulation.  For a
normal mass hierarchy, a five year run can detect $\ssttot$ at
3$\sigma$ for values as low as 0.003.  If the mass hierarchy is
inverted, the beam can be run in antineutrino mode, with a similar
signal-to-noise ratio, and event rate 55\% as high as for the neutrino
mode.

In an interesting coincidence for the 9300 km baseline, the second peak
of maximum $\nu_\mu$ disappearance occurs at the same energy as
the peak for $\nu_e$ appearance.  Thus, interesting measurements of
the parameters of $\nu_\mu$ disappearance may also be possible.

Phase 2 of the JHF-Kamioka proposal assumes the construction of the
Hyper-K detector with 40 times the fiducial mass of the Super-K detector.
Having this detector serve as the target for beams from both JHF and
Fermilab is a powerful way of placing constraints in the $\ssttot$-$\delta$
plane.  We have investigated values of $\ssttot$ ranging from 0.001 to
0.05, and find that an 8 year run can establish maximal CP violation at
3$\sigma$ or better throughout this range.  If the JHF antineutrino beam
has poor performance, using Fermilab and JHF neutrino beams alone
is almost as effective in constraining $\delta$.

\section{Acknowledgments}
\label{sec:ack}

We thank the NuMI collaboration for the use of their fast beamline
simulation.  In particular, we thank Debbie Harris for help with the use
of this software.

\end{document}